\documentclass{aa}  
\usepackage[dvipsnames]{xcolor} 
\usepackage{graphicx}
\usepackage{txfonts}
\usepackage{xcolor}
\usepackage{mathrsfs}
\usepackage{physics}
\usepackage{nicefrac}
\usepackage[normalem]{ulem}
\usepackage{soul}
\usepackage[colorlinks=true, citecolor=blue]{hyperref}
\usepackage{comment}

\newcommand{\FargoCPT}{\textsc{FargoCPT}\xspace}

\newcommand{\NIRVANA}{\textsc{Nirvana-iii}\xspace}
\newcommand{\vr}{\vec{r}}

\newcommand{\es}{\vec{e}_s}

\newcommand{\Nabla}{\vec{\nabla}}
\newcommand{\fourier}{\mathcal{F}}

\newcommand{\Hab}{H_{ab}}
\newcommand{\dab}{d_{ab}}
\newcommand{\dsqbytwo}{\frac{\dab^2}{2}}
\newcommand{\K}{\text{K}_{ab}}
\newcommand{\TQ}{Q_{\text{bi-fluid}}}

\begin{document}

\title{Self-gravity in thin protoplanetary discs:}
\subtitle{1. The smoothing-length approximation versus the exact self-gravity Kernel}

\author{S. Rendon Restrepo \inst{1}
\thanks{\email{srendon@aip.de}}
\and
T. Rometsch \inst{2}
\and
U. Ziegler \inst{1}
\and
O. Gressel \inst{1}
}
\institute{Leibniz-Institut für Astrophysik Potsdam (AIP), An der Sternwarte 16, 14482 Potsdam, Germany
\and
Institut für Theoretische Astrophysik, Zentrum für Astronomie (ZAH), Universität Heidelberg, Albert-Ueberle-Str. 2, 69120 Heidelberg, Germany}

 
\abstract
{Planet-forming discs are increasingly found to harbour internal structures such as spiral arms. The origin and evolution of these is often associated with the disc's own gravitational force. When investigating discs using a 2D approximation, it is common to employ an ad-hoc softening prescription for self-gravity. 
However, this approach ignores how the vertical structure of the disc is affected by the mass distribution of gas and dust. 
More significantly, it suppresses the Newtonian nature of gravity at short distances and does not respect Newton's third law.
}
{For overcoming the inherent issues associated with approximate descriptions, for instance, a Plummer potential, we aim to derive an exact self-gravity kernel designed for hydrostatically supported thin discs, which moreover incorporates a potential dust fluid component embedded in the gas.
}
{We develop an analytical framework to derive an exact 2D self-gravity prescription suitable for modelling thin discs. 
The validity and consistency of the proposed kernel is then supported by analytical benchmarks and 2D/3D numerical tests. 
}
{We derive the exact 2D self-gravity kernel valid for Gaussian-stratified thin discs. 
This kernel is built upon exponentially scaled modified Bessel functions and simultaneously adheres to all the expected features of Newtonian gravitation -- including point-wise symmetry, a smooth transition from light to massive discs as well as a singularity for vanishing distances, among others. 
Quite remarkably, the kernel displays a purely 2D nature at short distances, while transitioning to a fully 3D behaviour at longer distances. 
In contrast to other prescriptions found in the literature, it proves capable of leading to an additional, and previously unnoticed, source of gravitational runaway discernible only at infinitesimal distances. 
We finally note that our new prescription remains compatible with methods based on the fast Fourier transform, affording superior computational efficiency.
}
{ 
Our exact kernel formulation overcomes substantial limitations inherent in the smoothing-length approach. It permits a novel, fully consistent treatment of self-gravity in Gaussian-stratified thin discs.
The approach, that makes the usage of the Plummer potential obsolete, will prove useful to studying all common planet formation scenarios, which are often backed by 2D-flat numerical simulations.
Accordingly, in an accompanying paper, we will investigate how the occurence of the gravitational instability is affected.
}


\maketitle
%

\section{Introduction}

With the current advancements in numerical capabilities and optimised codes, conducting complex simulations of physical processes within protoplanetary discs (PPDs) has become routine.
Despite the increased computational power, high-resolution 3D simulations present practical challenges such as data storage, analysis or visualisation that remain unresolved.
More significantly, they remain computationally challenging, necessitating a resolution trade-off across the three spatial directions.
As a result, when studying discs numerically, one of the common approaches involves employing a 2D approximation, which neglects the vertical dimension and hence offers a significant gain of resolution in the $(r, \varphi)$ plane.
In particular, this approximation allows to conduct sufficiently resolved global simulations including the gravitational force of the disc exerted onto itself, commonly referred as self-gravity (SG).

Deriving the ``correct'' approximation for the effective gravity in a 2D disc is a subtle and tedious problem, that requires to define first what is meant by 2D approximation.
As an extreme case, this could mean living in a ``two-dimensional universe'', which implies that the laws of Nature are fundamentally different from our 3D space -- that is, the gravitational field generated by a point mass scales as $\propto 1/r$ \citep{1996_Landy}.
Of course, this does not apply in the case of discs, as they are objects embedded within a three dimensional space.
But based on theoretical expectations, and supported by PPD observations \citep{2016_pinte}, cataclysmic variable accretion discs \citep{2016_baptista} and galactic discs \citep{2014_hu,2019_vandokkum}, we expect discs to have low geometrical aspect ratios.
This justifies the flat, or razor-thin, disc assumption for the density vertical profile, which adopts a Dirac $\delta$ distribution.
In the particular case of axisymmetric discs, this assumption permits to derive the gravitational potential in the equatorial plane. 
For instance, this could be achieved by means of a continuous overlap of infinitely flattened homoeoid shells \citep[Sects. 2.5.1 and 2.6]{2008_binney_tremaine} or through direct integration, resulting in a closed form for the associated potential \citep{1953_durand, 2007_hure_hersant, 2008_hure_hersant}.
Although almost flat, PPDs are inherently three-dimensional structures, yet due to computational limitations or for the sake of analytical simplicity, the oversimplified Dirac profile assumption is often made.
To eliminate this limitation, the other approach, that we endorse in this paper, consists on averaging vertically the 3D SG force in a thin disc \citep{shengtai_2009,muller_kley_2012,2023_rendon_restrepo}.
As a result the obtained 2D force, meant to act in the midplane, encapsulates in average all information from the vertical structure of the disc. 
We emphasise that the thin disc structure originates from the vertical hydrostatic equilibrium, which, in the simplest case, leads to a Gaussian profile \citep{2022_armitage}.

For accounting the disc's vertical thickness in the 2D SG calculation and to presumably avoid numerical singularities, the gravitational potential of a thin disc was often approximated by a Plummer potential \footnote{Note that for the rest of the paper we adopt $G=1$.}, that is,
\begin{equation}
\Psi_{\rm Plumm}(\vr) = - \displaystyle \iint\limits_{\rm disc} \frac{ \Sigma(\vr')}{\sqrt{ |\vr-\vr'|^2 + \epsilon^2}} \, d^2 \, \vr'
\end{equation}
where the smoothing length (SL), denoted as $\epsilon$, was regarded during a considerable period of time as a free parameter.
Over the years, several values have been proposed and theoretical findings converged towards a softening length proportional to the disc's scale height \citep{2009_hure_pierens, 2011_hure_hersant, 2015_hure_trova}.
The prevailing value frequently used in numerical simulations being $\epsilon_g/H_g = 0.6-1.2$ \citep{muller_kley_2012}.
We highlight that \citet{shengtai_2009} identified another prescription that does not utilize a softening prescription.
However, it has received little attention, likely due to its apparent complexity and the lack of efficient, compatible numerical methods. 
In the latest work to date by \citet{2023_rendon_restrepo}, it was shown that while this widely used prescription models correctly the long range SG interaction, it tends to underestimate the mid/short range interaction by up to 100\% -- a discrepancy that aligns with the removal of the Newtonian behaviour in the presence of softening \citep{1989_adams, hockney2021computer, 2008_baruteau, 2015_young_clarke}.
As a consequence, these authors introduced a space-varying smoothing length (SVSL), departing from the simple constant SL approach.
This SVSL allows the 2D force to better fit the vertically averaged 3D counterpart.
Further, they generalised their correction to bi-fluids, that is, the case when a dust phase is embedded in the gas. They found that two additional SLs are required for accounting the gravitational interactions of dust with dust and the crosswise interaction of dust with gas.
This could be of great importance, since the low levels of turbulence in the vertical direction -- supported both by observations of thin dust layers \citep{2020_villenave, 2022_villenave}, and by 3D shearing box simulations \citep{2021_baehr} -- may suggest that dust SG could be comparable to the one of gas at small scales.

Even if \citet{2023_rendon_restrepo} solved the main issue with their new prescription, they nonetheless used a number of mathematical simplifications which, are debatable.
In particular, they assumed that the relative scale heights between component $a$ and $b$, expressed via $\eta_{ab}=H_a(\vr')/H_b(\vr)$ are constant over the 2D disc. This, however, breaks the $\vr/\vr'$ symmetry in the SG force.
This symmetry breaking results in a spurious radial acceleration, which needs to be compensated \citep{baruteauPredictiveScenariosPlanetary2008a, 2024_rometsch}.
Furthermore, the authors disregarded the effect of SG on the vertical direction, which could significantly affect the layering of the disc \citep{1999_Bertin_Lodato,lodato_2007} and thus also indirectly the 2D SG force.
As a consequence, \citet{2023_rendon_restrepo} prescription is only valid for comparatively light discs, whose Toomre's parameter satisfies $Q_g \gtrsim 5$, and cannot be used self-consistently in simulations of Gravitational Instability (GI) where Toomre's parameter is expected to reach values smaller than unity \citep{2001_gammie, 2016_takahashi, 2017_baehr, 2021_bethune}.
In a broader context, this could profoundly impact a broad range of planet formation scenarios, since, at some point or another, the SG force should come into play in maintaining the gravitational cohesion of the formed objects \citep{2023_rendon_gressel_PPVII_poster}.
Finally, the work of \citet{2023_rendon_restrepo} should be considered as a preliminary step in view of the dust stratification, since it was presumed to be Gaussian without providing further details. 
This can be readily contradicted for strong SG of dust where the sedimentation follows a $\sech^2$-profile \citep{2020_klahr_schreiber, 2021_klahr_schreiber}.
Furthermore, the mentioned assumption disregards the gravitational contribution of the gas to the vertical layering.
We conclude that all issues and ambiguities raised in this paragraph need to be corrected in order to conduct consistent 2D simulations of single or bi-fluids incorporating SG.

In this first paper of a series of two, we aim to provide an exact SG kernel that must be used in 2D global simulations of discs -- a formulation that is valid from light to massive discs all at once.
In the second accompanying paper we study how this new kernel affects the GI planet formation scenario. 
We begin by recalling briefly the vertical stratification of self-gravitating, bi-fluid discs and provide the exact SG kernel in Section~\ref{Sect: Self-gravity for two phase discs}.
Then, in Section~\ref{sec: advantages associated with the Bessel kernel}, we highlight the relevance of this kernel and discuss how it possibly leads to a new runaway situation.
In Section~\ref{Sect: validation}, we benchmark our kernel against well established analytical results from thin disk literature and 2D/3D numerical tests.
Finally, in Section~\ref{Sect: discussion} we propose a discussion on numerical methods, the limitations of our approach, the indirect term and the consistency of 2D Poisson's equation.
Important equations and derivations can be found in the appendices.

\section{Self-gravity for two-phase discs}\label{Sect: Self-gravity for two phase discs}

%
\begin{table}
\caption{Definitions and list of abbreviations}             
\label{tab:list abbreviations}      
\centering          
\begin{tabular}{l l l}     
\hline\hline       
Abbrev.             & Definition+Name                   & Symbol                                         \\ 
\hline              
SG                  & Self-gravity                      &                                                \\ 
SL                  & Smoothing length                  & $\epsilon_g \propto H_g$                       \\
SVSL                & Spatially varying SL    &                                                \\
                    & Mutual distance/separation     & $s = ||\vr-\vr'||$                             \\
                    & r.m.s scale height                & $\Hab^{sg}=\sqrt{\frac{H_a^{sg}+H_b^{sg}}{2}}$ \\
                    & Normalised distance               & $d_{ab} = s/\Hab^{sg}$                         \\
                    & Self-gravity force kernel         & $\K$ (Eq.~\ref{Eq: SG force kernel exact})     \\
                    & Dust-to-gas scale height          & $\eta$ (Eq.~\ref{Eq: def eta})                 \\
\hline                  
\end{tabular}
\end{table}

Estimating in-plane SG forces in thin discs involves a two-step process. 
The initial step, often overlooked, involves determining the vertical hydrostatic equilibrium of the system. 
The resulting stratification is subsequently used, in the final phase, as an input for vertically integrating all forces -- including, in our specific case, SG.

Accordingly, in next sections we recall the results of \citet{2025_rendon_restrepo_et_al} regarding the stratification of a bi-fluid disc and the key steps taken by \citet{2023_rendon_restrepo} for computing SG in 2D, bi-fluid simulations of PPDs. 
Both results combined enable us to propose an exact SG kernel that works for a broad range of Toomre's parameter, as well as a bi-fluid. At the same time, it frees us from the problems inherent of a Plummer potential.
To enhance readability of this paper, we have compiled the definitions and abbreviations of main
quantities in Table~\ref{tab:list abbreviations}.

\subsection{The thin self-gravitating, bi-fluid disc}\label{sec: the thin self-gravitating, bi-fluid disc}

For simplicity, it is often assumed that the vertical structure of the gas and dust phases composing a PPD is set either by the balance between the star's gravity and by pressure, or by viscous stirring, respectively.
However, the vertical component of the disc's SG should be included in this equilibrium and particularly when modelling class 0/I disks or the outskirts of PPDs where the SG force is predominant over the star's gravity.
This is most likely the case in the Elias~2-27 system \citep{2016_perez, 2022_parker} and in IM~Lupi and GM~Aur \citep{2023_lodato}.

When the gravitational contribution of gas and dust are treated separately in massive discs, their respective layering are no longer Gaussian, but are set by a $\sech^2$ function \citep{1942_spitzer, 2020_klahr_schreiber,2021_klahr_schreiber}.
Additionally, when the disc is solely composed of gas, it is possible to connect the stratification of light and massive discs using a ``biased'' Gaussian stratification. Such an approach incorporates all SG information into a modified scale height dependent on Toomre's parameter, as shown by \citet{1999_Bertin_Lodato}.
\citet{2025_rendon_restrepo_et_al} extended their work and showed that it is not possible to disentangle the contribution of gas and dust to SG, and proposed a very accurate and general solution valid for any SG strength, of gas and/or dust.
At first order, their solution is:
\begin{equation}\label{Eq: stratification gas and dust}
\left\{
\begin{array}{cc}
\rho_g(\vr,z) & = \displaystyle \frac{\Sigma_g}{\sqrt{2\pi} H_{g}^{sg}} \exp\left[-\frac{1}{2} \left(z/H_{g}^{sg}\right)^2\right] \\ [8pt]
\rho_d(\vr,z) & = \displaystyle \frac{\Sigma_d}{\sqrt{2\pi} H_{d}^{sg}} \exp\left[-\frac{1}{2} \left(z/H_{d}^{sg}\right)^2\right] 
\end{array}
\right.
\end{equation}
with:
\begin{equation}\label{Eq: stratification parameters}
\left\{
\begin{array}{lll}
H_g^{sg} &=& \displaystyle \sqrt{\frac{2}{\pi}} H_g f(\TQ)   \\
H_d^{sg} &=& \displaystyle \sqrt{\frac{2}{\pi}} H_d f(\TQ) \\
\TQ        & =  & \left( \frac{1}{Q_g} + \frac{1}{Q_d} \right)^{-1} \\
f(x)     & =  &\displaystyle \frac{\pi}{4 x} \left[ \sqrt{1+\frac{8 x^2}{\pi}} -1 \right]
\end{array}
\right.
\end{equation}
where $H_g$ is the thermal gas scale height and $H_d$ is the dust diffusive scale height.
The quantities $Q_g=\frac{c_g \Omega_K}{\pi G \Sigma_g}$ and $Q_d=\frac{c_{d} \Omega_K}{\pi G \Sigma_d}$ are the Toomre's parameter of gas and dust, respectively.
The effective dust sound speed is $c_d$.
Meanwhile, the term $f(\TQ)$, present in the definition of both scale heights, indicates that both fluids experience an equal gravitational influence from the star and from their combined mass distributions.
Indeed, the crosswise gravity of both fluids is encapsulated by the general Toomre's parameter, $\TQ$, defined as the harmonic average of its constituents.
For a detailed discussion, we invite readers to refer to \citep{2025_rendon_restrepo_et_al}.
Contrary to the razor-thin disc assumption mentioned in the introduction, the approach discussed in this section is able to accommodate any vertical stratification for the gas and dust phases.
For instance, in the case of uniquely dust, it could be applicable to systems with limited sedimentation \citep{2023_lin} to high settling \citep{2020_villenave, 2022_villenave}.

As we will demonstrate in the following section, the vertical profiles exhibited in Eq.~\ref{Eq: stratification gas and dust} can be elegantly exploited for obtaining the exact 2D SG force via a kernel description.

\subsection{Exact self-gravity kernel}\label{sec: Exact self-gravity force correction}

\begin{figure}
\centering
\includegraphics[width=\hsize]{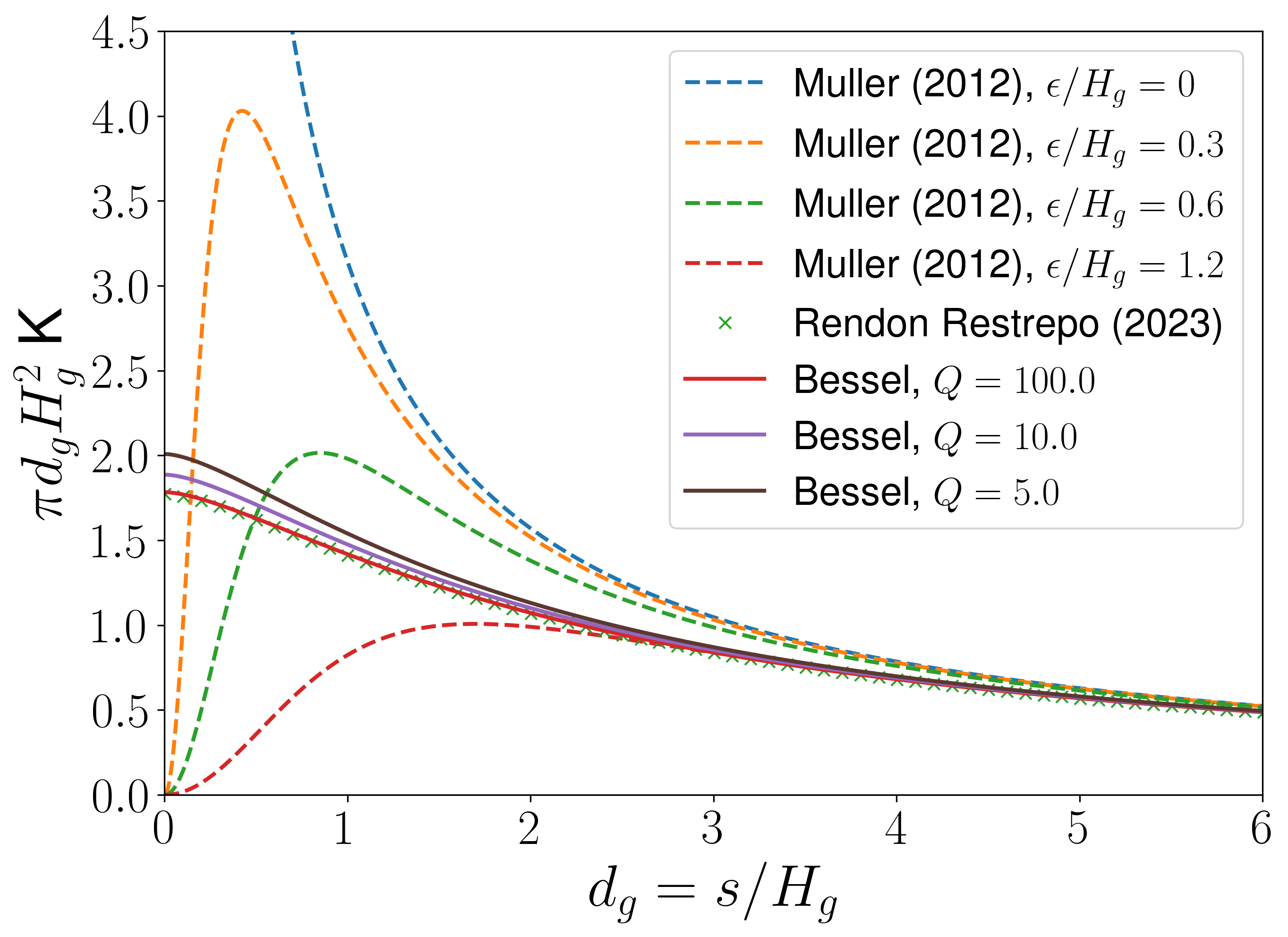}
\includegraphics[width=\hsize]{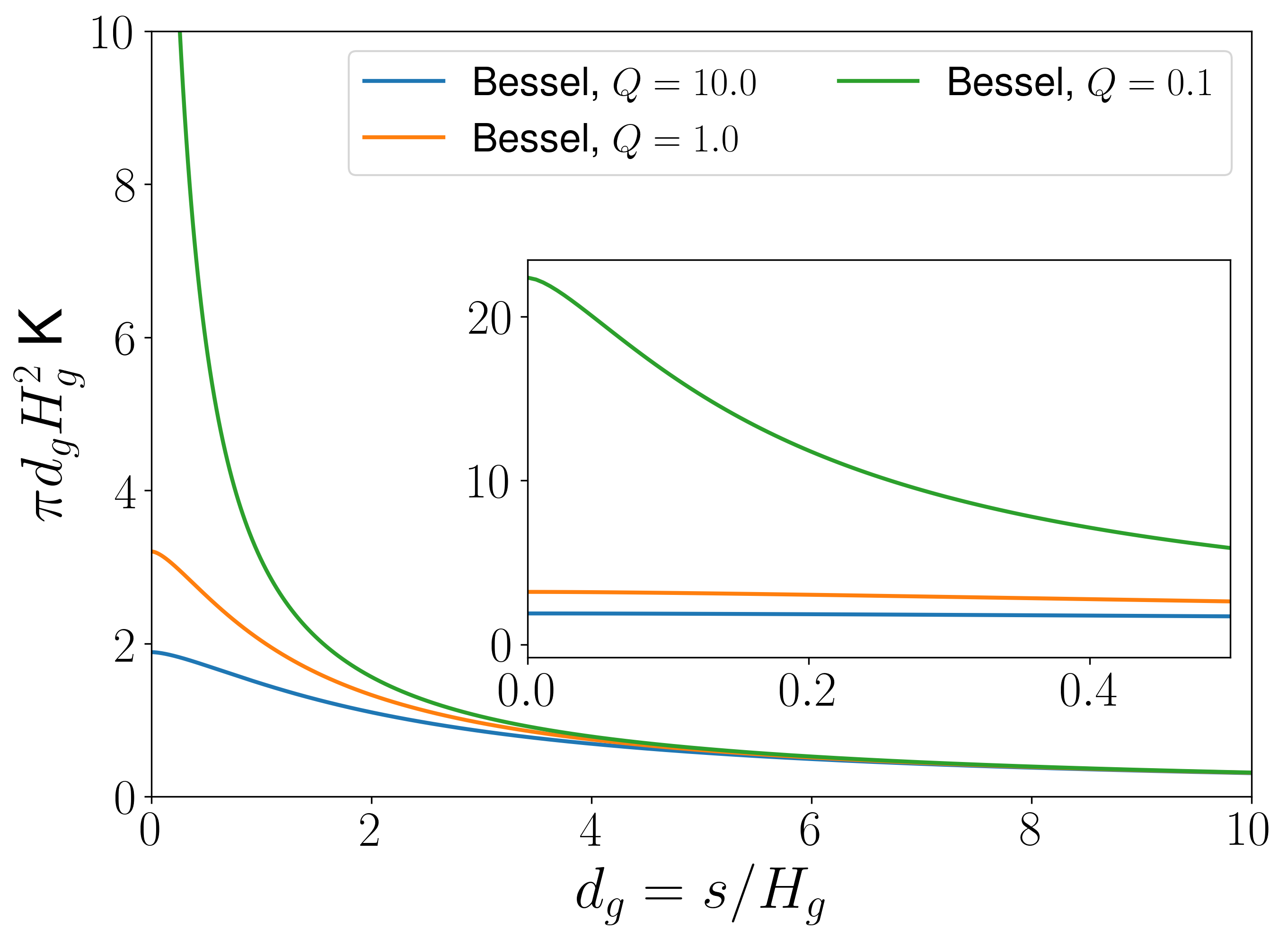}

\caption{Normalised self-gravity kernels, $\pi d_g H_g^2 \, \text{K}$, for light (top) and massive (bottom) discs. 
For light discs, the prescriptions of \citet{muller_kley_2012} and \citet{2023_rendon_restrepo} are insensitive to the Toomre's parameter.
This is in contrast with the Bessel prescription, which scale as $1/Q_g$ in the limit of massive discs.
When a constant SL prescription is used, the kernel vanishes at short distances, whereas for a vanishing SL, the kernel diverges, suggesting an overestimation of SG.
The Bessel prescription is the unique prescription that is simultaneously inversely proportional to $Q_g$ for massive discs, does not vanish at the singularity and permits a smooth transition between massive and light discs.
} 
\label{fig: force correction Bessel}
\end{figure}

We consider two fluids (gas and dust) with the Gaussian vertical stratification defined by Eqs.~\ref{Eq: stratification gas and dust}-\ref{Eq: stratification parameters}.
In 3D, the SG force per unit volume exerted by the PPD on gas and dust parcels are expressed as:
\begin{equation}\label{Eq:3D volume force total}
\begin{array}{lll}
\vec{f}^{g,tot}_{3D}(\vr,z) &=& -\rho_g(\vr,z) \left[ \Nabla \Phi_g + \Nabla \Phi_d \right], \\ [4pt]
\vec{f}^{d,tot}_{3D}(\vr,z) &=& -\rho_d(\vr,z) \left[ \Nabla \Phi_g + \Nabla \Phi_d \right], 
\end{array}
\end{equation}
where $\Phi_g$ and $\Phi_d$ are the distinct gravitational potentials associated with the gas and dust discs, respectively.
For the sake of generality and conciseness, we will denote the two fluid phases as `a' and `b' for the remainder of the paper.
The 2D SG force per unit volume exerted by the disc made of fluid `a' on a volume element of fluid `b' is (see Eq. 4 in \citep{2023_rendon_restrepo}):
\begin{equation}
\begin{array}{lll}\label{Eq: SG force general}
\vec{f}^{a \rightarrow b}_{2D}(\vr) 
             & = &\displaystyle - \rho_{b}(\vr,z=0) 
                   \iint\limits_{disc} 
                   \,\rho_{a}(\vr',z=0) \, s \, \es  \\
             &   & \displaystyle \left( \, \, \iint\limits_{z, z'=-\infty}^{+\infty} 
                   \displaystyle \frac{e^{-\frac{1}{2} \left(z/H_{b}^{sg}(\vr)\right)^2} e^{-\frac{1}{2} \left(z'/H_{a}^{sg}(\vr')\right)^2}}{\left(s^2+(z-z')^2\right)^{{3}/{2}}} \, dz \, dz' \right) d^2\vr'
\end{array}
\end{equation}
where  $s=||\vr-\vr'||$ stands for the separation between two fluid elements and $\es=\left(\vr-\vr'\right)\big/s$ is a unit vector in the 2D plane.
We highlight that all variables linked to phase `a' and `b' are dependent on $\vr'$ and $\vr$ positions, respectively.
Hence, for the rest of the paper we skip all explicit dependencies on position, except when a distinction is necessary. 
We highlight that for a disc made of gas and dust, our naming convention leads to four possible combinations $(a,b)=\left\{ \right.$(gas, gas), (gas, dust), (dust, dust), (dust, gas) $\left. \right\}$ for computing $\vec{f}^{a \rightarrow b}_{2D}$.
With these clarifications, Eq.~\ref{Eq: SG force general} can be rearranged in the following manner:
\begin{equation}\label{Eq: volume force 2D}
\vec{f}^{a \rightarrow b}_{2D}(\vr) 
             = - \Sigma_{b} 
               \iint\limits_{disc} 
               \, \Sigma_{a} \, \K \, \es \, d^2\vr'
\end{equation}
where:
\begin{equation}\label{Eq: SG force kernel general}
\begin{array}{lll}
\K
      & = & \displaystyle \frac{1}{2 \pi} \frac{s}{H_a^{sg} H_b^{sg}} 
            \displaystyle  \, \, \iint\limits_{z, z'=-\infty}^{+\infty} 
            \displaystyle \frac{e^{-\frac{1}{2} \left(z/H_b^{sg}\right)^2} e^{-\frac{1}{2} \left(z'/H_a^{sg}\right)^2}}{\left(s^2+(z-z')^2\right)^{{3}/{2}}} \, dz \, dz'  \\
\end{array}
\end{equation}
is the SG force kernel between fluids `a' and `b'
\footnote{We can also write the SG kernel as $\K = {L^{a b}(\dab)}/(\pi \Hab^{sg} \, s)$, where $L^{a b}$ is the self-gravity force correction (SGFC) defined in \citet{2023_rendon_restrepo}. For clarity we will not refer to the SGFC here.}.
The column densities are $\Sigma_i$.
Previously, \citet{2023_rendon_restrepo} attempted to provide an accurate approximation of the integral defined by Eq.~\ref{Eq: SG force kernel general} employing a SVSL, which, despite its accuracy, remains burdensome. In particular, it does not respect the $\vr/\vr'$ symmetry and is not valid in the regime where SG starts to become significant (i.e $Q \lesssim 5$).
Luckily, when the vertical component of SG is disregarded, the double integral defining the SG force kernel can be assessed analytically \citep[Eqs. 20-22]{shengtai_2009} leading us to the derivation of the force kernel in the general case of a self-gravitating bi-fluid as seen below (see appendices~\ref{app:useful function} and \ref{app: integration} for a detailed derivation):
\begin{equation}\label{Eq: SG force kernel exact}
\begin{array}{lll}
\K
      & = & \displaystyle \frac{1}{\sqrt{\pi}}\left( \Hab^{sg} \right)^{-2} \frac{\dab}{8} 
            \displaystyle \exp\left(\frac{\dab^2}{8} \right) \left[ K_1\left(\frac{\dab^2}{8} \right) - K_0\left(\frac{\dab^2}{8} \right) \right]
\end{array}
\end{equation}
where $K_\alpha$ are modified Bessel functions of the second kind and order $\alpha$.
The notation $d_{ab}=s/\Hab^{sg}$ stands for the normalised distance between two fluid elements.
The root mean square of $H_a^{sg}$ and $H_b^{sg}$ is defined by:
\begin{equation}\label{Eq: mean root square height}
\Hab^{sg} = \sqrt{\frac{{H_a^{sg}}^2 + {H_b^{sg}}^2}{2}}
\,.\end{equation}
The above-mentioned distance $\Hab$ not only combines the different fluids contribution to the scale height but also takes into account the different disc positions $\vr$ and $\vr'$.
We highlight, that the unique definition of above r.m.s scale height (Eq.~\ref{Eq: mean root square height}) permits to lift the $\vr/\vr'$ symmetry issues associated to a 2D approximation, even when using the Plummer potential prescription. The symmetry issues are avoided because Newton's third law is fulfilled as long as the force between two density columns is symmetric.

The closed form expressed in Eq.~\ref{Eq: SG force kernel exact} offers several advantages over the one proposed by \citet{2023_rendon_restrepo}.
Despite employing special mathematical functions, this formula remains both simple and general. 
Notably, there is no longer a necessity for the auxiliary functions $(\lambda, \delta)$ introduced when using a SVSL.
Furthermore, Eq.~\ref{Eq: SG force kernel exact} encompasses all mutual interactions between different phases embedded in a PPD.
Finally, this new kernel formulation achieves the $\vr/\vr'$ position symmetry, implying Newton's third law, as well as the interchangeability symmetry between fluid elements `a' and `b'.
All these aspects reassure us in the accurate representation of the 2D SG when employing our new kernel.
We stress, that our work differs from \citet{shengtai_2009} due to the incorporation of the bi-fluid analysis (readily adaptable to N-fluids, see Sect. \ref{sec: efficient numerical method} for a discussion) and our consideration of how the SG of both components affects their vertical density profile.
Indeed, in our investigation, the vertical component of SG is encapsulated in the definition of the revised scale height (Eq.~\ref{Eq: stratification parameters}). 
We stress that these two last aspects are novel. 
The chosen approach makes our method similar to a 1D + 2D problem, justified by the fact that the timescales of vertical settling are much smaller than the in-plane dynamic timescales.
Further, we want to highlight that due to computational costs, \citet{shengtai_2009} did not encourage to use the Bessel kernel but instead an interpolated formula.
Presumably, they did not contemplate that Fast Fourier Transform (FFT) methods were still valid for such prescription and that, under certain circumstances, the evaluation of the Bessel kernel is only done once for a whole simulation.
This is regrettable because there are a lot of physical lessons to be drawn from their formula, which we we will explore in the next section.

\section{Advantages associated with the Bessel kernel}\label{sec: advantages associated with the Bessel kernel}

In this section, we explore the benefits of putting our prescription to use.
To achieve this, we examine how it compares to other prescriptions from the literature, assess how SG is rescaled when considering a dust phase, and demonstrate that our kernel is the only one enabling a potential runaway process at infinitesimal distances.

\subsection{Comparison with other kernels from the literature}\label{sec: comparison with other kernels}

In order to show the interest of our findings, we depicted in Fig.~\ref{fig: force correction Bessel} a comparison between our Bessel kernel and the prescriptions suggested by \citet{muller_kley_2012} and \citet{2023_rendon_restrepo}.
We remind that the first approach relies on a constant smoothing length, which we set at $\epsilon/H_g \in [0, 0.3, 1.2]$, three customary values in several numerical studies \citep{2011_paardekooper, 2015_young_clarke, 2016_zhu, baruteau_2016, 2018_vorobyov}.
The second approach is intended for weak SG conditions, $Q_g\gtrsim 5$, and employs a space-varying smoothing length.
For simplicity, this comparative study assumes a gas-only disc.
To ensure finite values at the singularity ($d_g=0$) we normalised the different kernels by $\pi^{-1} d_g^{-1} H_g^{-2}$.
In particular, the last factor allows us to account for modifications in the scale height due to SG whenever it is incorporated in the kernel's prescription—as is the case only for our Bessel kernel.
In top panel of Fig.~\ref{fig: force correction Bessel}, we compare the kernels designed for application in the weak SG regime, while the bottom panel shows the Bessel kernel applied to massive discs.

As expected, our plots revealed similar behaviour among all kernels for the long-range interaction: $d_g \gtrsim 3$ when the disc is light or $d_g \gtrsim 3 Q_g$ when the disc is massive.
Notably, all kernels scale as $\propto d_g^{-2}$ for large distances, which respects the Newtonian behaviour.
Conversely, all these prescriptions reveal distinct behaviours at short range, which we will discuss next.
First, we highlight that for a vanishing SL the kernel adopts an inverse square law at small distances, which highly overestimates SG. 
For the constant and non-vanishing SL, two scenarios arise.
When the SL approaches zero (e.g., $\epsilon/H_g=0.3$), the SG vanishes for small distances but is overestimated for $d_g \in [0.2, 3]$.
Conversely, larger SL (e.g., $\epsilon/H_g=1.2$) necessarily underestimate SG for $d_g \in [0.2, 3]$.
This last limitation was corrected by \citet{2023_rendon_restrepo} prescription, as their normalised kernel does not indeed vanish at the singularity, aligning more closely with expectations for gravitational interactions: the closer two bodies are, the stronger their mutual gravitational pull, a fundamental principle that must remain true even in the 2D approximation.
However, this latter kernel is insensitive to the Toomre's parameter of the disk, which is inconsistent.
The Bessel kernel proposed in this work addresses this issue, as its value scales inversely with the Toomre's parameter (bottom panel of Fig.~\ref{fig: force correction Bessel}), matching the expectation that SG strengthens in massive discs.
Furthermore, it does not vanish at the singularity, a feature consistent with the nature of gravity.
Finally, the Bessel kernel permits a smooth transition between light ($Q_g \gtrsim 5$) and massive discs ($Q_g\lesssim1$).
These are expected features from a coherent 2D SG kernel. 
We also note that with the Bessel kernel, the effects of SG start to become appreciable when $Q_g \lesssim 5$.
It is noticeable that for the short and long range interactions, the Bessel kernel scales as $\propto d_g^{-1}$ and $\propto d_g^{-2}$, respectively, in agreement with an apparent 2D-3D gravity that we discuss next. 
Finally, we highlight that the point-wise symmetry of our Bessel kernel, enabled by the root mean square scale height definition (Eq. \ref{Eq: mean root square height}), eliminates the need for correction terms arising from radial self-accelerations \citep{baruteauPredictiveScenariosPlanetary2008a}.

\subsection{The gravitational character of our formalism}\label{sec: nature of gravity ?}

It is worth mentioning a property that we find quite intriguing: while the long-range interaction of our kernel mirrors that of a gravitational force in a 3D space, at short range, it exhibits characteristics akin to a purely 2D nature since the force scales as $\propto 1/s$ (see the introduction).
This feature is particularly important for the short-range interaction of SG within the context of PPDs -- and it insightfully highlights the inherent discrepancies of the Plummer potential paradigm.

On the one hand, as the smoothing length approaches the disc scale height, $\epsilon \sim H_g$, it causes a significant underestimation of the short-range effects of SG, potentially hindering the formation of gravitationally bound objects. 
However, on the other hand, if the smoothing length vanishes, it is implied that the scaling of the 2D force is $1/s^2$, which in turn results in a substantial overestimation of SG.
Specifically, in the context of GI, we anticipate that the use of a vanishing SL could result in an overestimation of the formation of bound objects. 
This may offer a new perspective on the results reported by \citet{2018_vorobyov}, among others.
The realism of GI simulations employing a finite yet small SL, in works like \cite{2011_paardekooper} and \cite{2016_takahashi}, appears more challenging to predict. 
While the overestimation of SG in the range $d_g \in [0.2, 3]$ might suggest an amplification of GI, the suppression of SG at small scales could indicate the inability of fragments to further contract gravitationally, thereby promoting their destruction upon encounters with spirals. 
A more in-depth numerical investigation is required to evaluate the consequences of employing different gravity prescriptions.
This will be the focus of our upcoming paper.

\subsection{Gas and dust kernels comparison}

\begin{figure}[h]
\centering
\includegraphics[width=\hsize]{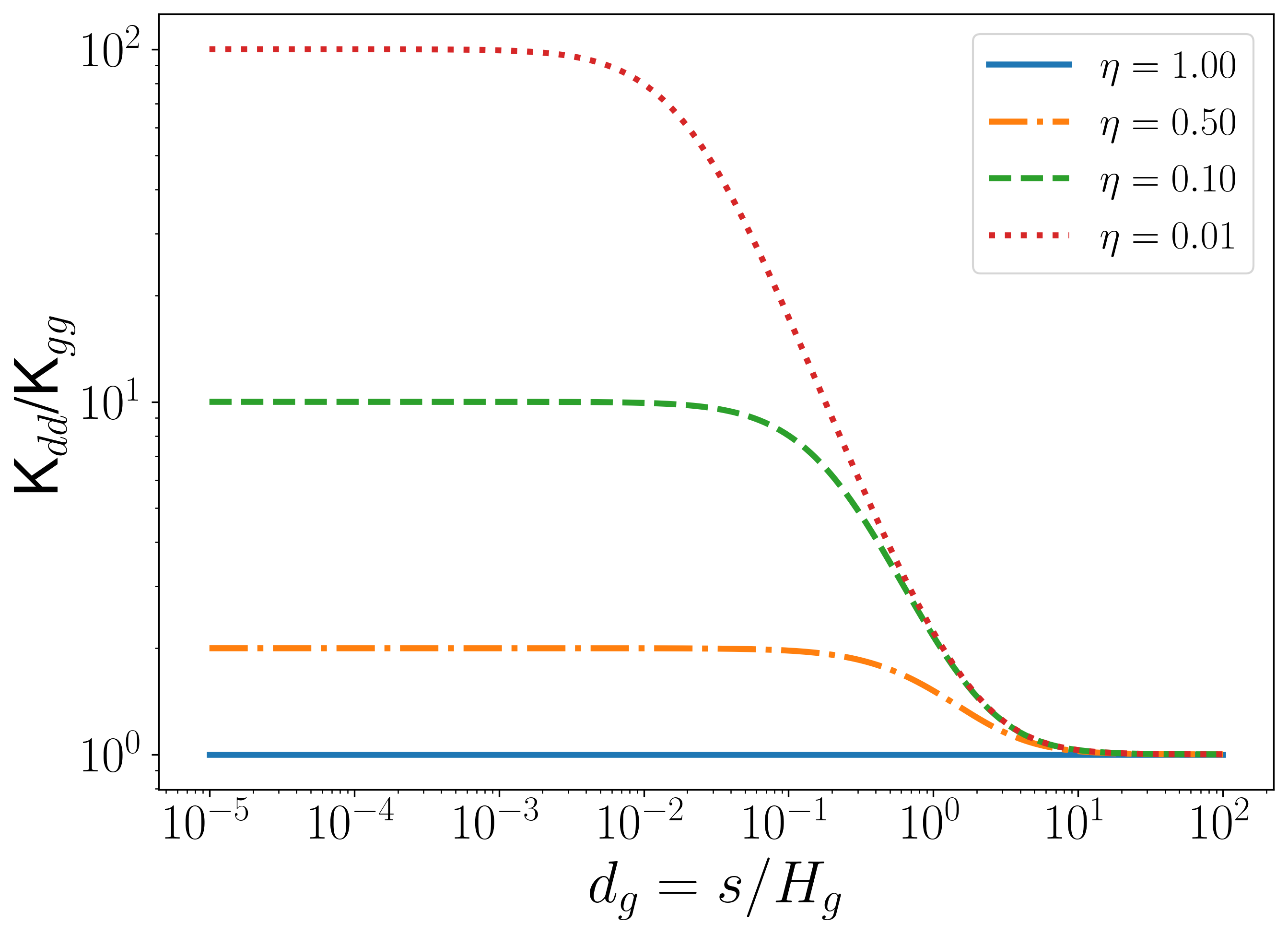}
\includegraphics[width=\hsize]{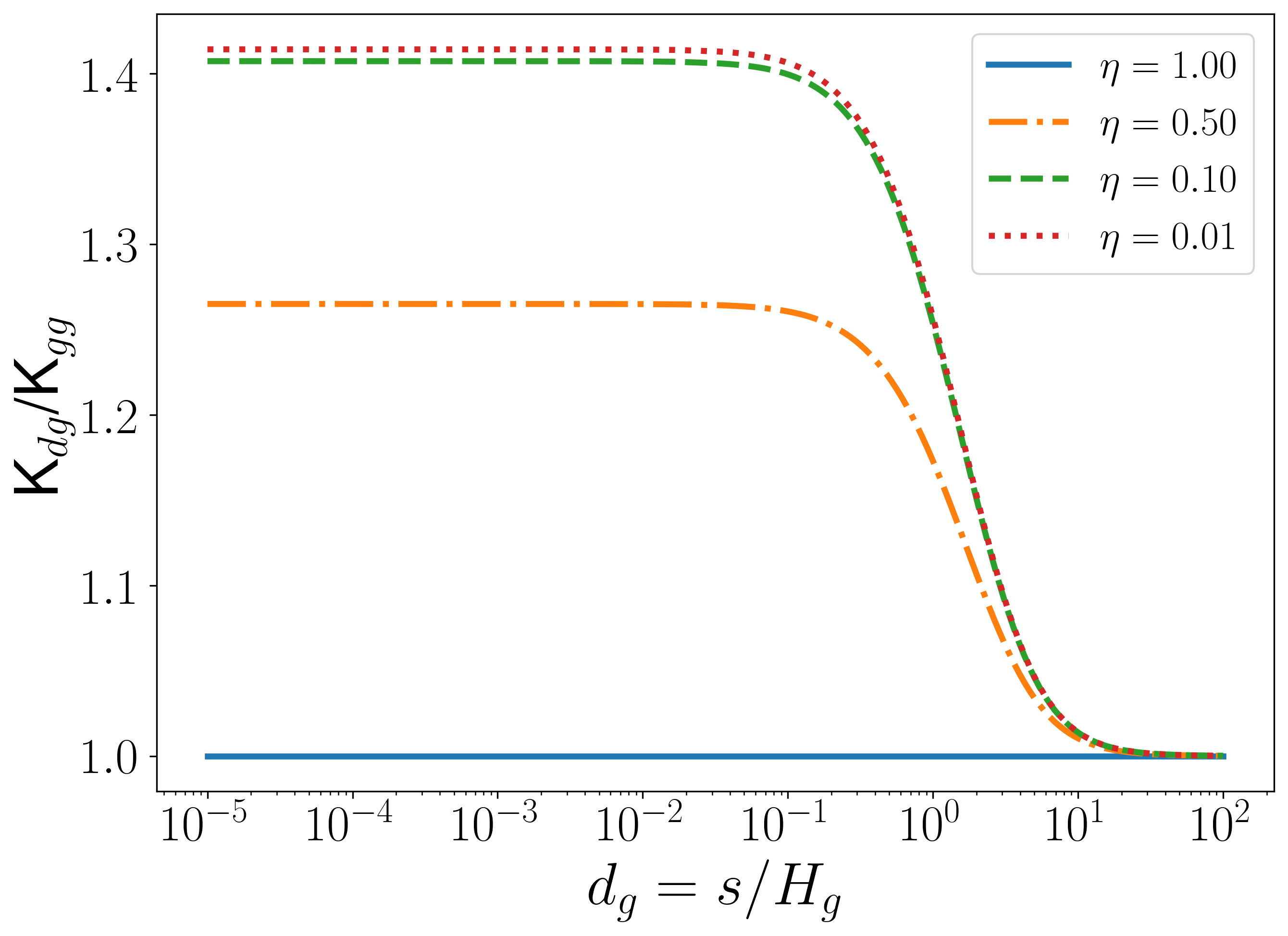}

\caption{Normalised kernels associated with dust with respect to distance for different dust-to-gas scale heights, $\eta$.
The dust-dust, $K_{dd}$, and dust-gas, $K_{dg}$, kernels are insensitive to $\eta$ at large distances, an aspect also in agreement with the smoothing length prescription (see Fig. 13 of \citep{muller_kley_2012}).
The dust-dust kernel scales with $1/\eta$ at short range, which suggests that the SG acceleration due to dust can be comparable of the one of gas even for $Z=\Sigma_d/\Sigma_g = 0.01$.
At short distances, the normalised dust-gas kernel converges quickly towards $\sqrt{2}$ (i.e., with respect to decreasing $\eta$).
As a consequence this kernel adopts a nature akin to the kernel associated with a planet-disc interaction.
} 
\label{fig: dust contribution vs gas contribution}
\end{figure}

An important feature of our analysis lies in the fact that our kernel does not apply solely to gas discs but it can also account for an embedded dust fluid.
Such situation results in four different contributions to the SG forces: one is attributed
to the gas disc, another to the dust layer and two to the mutual interaction between the gas and dust discs.
Even if the crosswise contributions are not commutative, above four contributions only necessitate the evaluation of three kernels: $K_{gg}$, $K_{dd}$ and $K_{gd}=K_{dg}$.
As a consequence, we aim to compare how the kernels involving dust, $K_{dd}$ and $K_{dg}$, compare to the kernel of a disc uniquely made of gas for different dust-to-gas scale heights ratios:
\begin{equation}\label{Eq: def eta}
\eta=H_{d}^{sg}/H_{g}^{sg}\,.
\end{equation}
In top panel of Fig.~\ref{fig: dust contribution vs gas contribution}, we depict the dust-to-dust kernel, while in the bottom panel we study the dust-to-gas kernel, which accounts for the crossed contribution.
Both quantities were normalised with respect to the gas-gas kernel, $K_{gg}$.
Motivated by highly settled discs \citep{2020_villenave, 2022_villenave} and also by younger systems with limited sedimentation \citep{2023_lin}, we constrained our investigation to $0.01 \leq \eta \leq 1$.
As expected, for both cases we observe that the kernel matches the one of a system only made of gas at long distances.
Indeed, at long separations a force SG kernel does not depend on the value of the scale height, a feature that also applies within the smoothing length prescription -- see Fig. 13 of \citet{muller_kley_2012}.
Most interestingly, we notice that at short distances, $d_g\lesssim \eta$, the dust-dust kernel is proportional to $1/\eta$.
For instance for $\eta=0.01$, this suggests that there is an enhancement of dust SG by a factor of 100.
Acknowledging that we expect column density dust-to-gas ratios of $Z=\Sigma_d/\Sigma_g=0.01$, this suggests that the SG acceleration due to dust is comparable, if not higher, than the one of gas in the limit of short distances.
On numerical grounds, this should be however taken with a grain of salt since currently there are little chances that such distances can be resolved by global simulations, except maybe when using adaptive mesh refinement techniques or in the context of shearing box simulations.
Regarding the normalised gas-dust kernel, we observe that, at short distances, it rises very quickly (with respect to decreasing $\eta$) towards a limiting value of $\sqrt{2}$.
Surprisingly, this means that the kernel tends towards the planet-disc kernel \citep[see Eq.~(20) of][]{muller_kley_2012}, a feature already pointed out by \citet{2023_rendon_restrepo}.
We will provide a physical explanation of this result in Sect.~\ref{Sect: validation}, below.

\subsection{Gravitational runaway}\label{sec: gravitational runaway}

In the following, we explore the ability of the Bessel kernel to potentially lead to a runaway process.
Let us consider a disc only made of gas where a perturbation caused a local and infinitesimal mass increase over an infinitesimal distance.
First, we focus in a massive disc ($Q_g \lesssim 1$), where the local mass increase will, in turn, decrease the local Toomre's parameter.
But in the specific case of massive discs the kernel at the singularity\footnote{Physically and in our context, the singularity is the region of space where a given fluid element is affected by its own gravity.} is $K_{gg} \sim \left(\sqrt{2} Q_g H_g s \right)^{-1}$.
Subsequently, the local SG is amplified, which feeds into the mechanism, resulting in a possible runaway process.
Now, consider the same scenario for a light disc, where the Toomre's parameter is large.
In such a case, the kernel at the singularity is independent of the density perturbation: $K_{gg} \sim \left(\sqrt{\pi} H_g s \right)^{-1}$.
Thus, this time, any local increase of density leaves the kernel unaffected, which does not amplify the mechanism and possibly shuts it down.
We believe that such a runaway process, which mimics a potential collapse in the vertical direction (through the reduction of the vertical scale height), is uniquely captured by our prescription.

We found that out of all prescriptions presented in Sect.~\ref{sec: comparison with other kernels}, only ours permits to maintain a gravitational collapse at infinitesimal distances.
Indeed, for \citet{muller_kley_2012} kernel, the rationale of the above paragraph would never lead to a runaway process at infinitely small scales since this kernel vanishes at the singularity, preventing any amplification of gravity caused by the kernel.
A runaway is only possible if the length of the perturbation is of the order $\sim H_g$ for the \citet{muller_kley_2012} kernel but this time the runaway is solely driven by the overdensity term present in the total force calculation ($\Sigma_g$ term in Eq.~\ref{Eq: volume force 2D}).
Similarly, for the SVSL formalism proposed by \citet{2023_rendon_restrepo}, even if the kernel does not vanish at the singularity, the runaway would be driven by the overdensity since the kernel is insensitive to the local Toomre's parameter increase.
All the aspects raised above will be addressed in more detail in the accompanying paper.
In following section, we propose to validate our kernel against analytical and numerical tests.

\section{Analytical and numerical validation}\label{Sect: validation}

We found it puzzling that the mathematical expression of the Bessel kernel (Eq.~\ref{Eq: SG force kernel exact}) is very similar to the kernel of a planet-disc interaction \citep[Eq. 20]{muller_kley_2012} by a scaling factor of $\sqrt{2}$ in the normalised distance.  
Convinced that this similarity is not coincidental, we proceed to demonstrate it within our formalism.
To elaborate, let's consider a disc made of two fluids: one gas component and a second ``fluid'' component that we consider as a planet.
From a purely gravitational perspective, a planet is nothing more than a fluid with a Dirac distribution in all space directions.
Accordingly, the ``fluid'' planet is a point and its density in the 2D plane can be expressed as:
\begin{equation}
\Sigma_p (\vr') = \frac{m_p}{\pi R_p^2}  \chi\left(\frac{||\vr'-\vr_{c,p}||}{2 R_p}\right)\,,
\end{equation}
where $\chi$ is the rectangular function, $R_p$ and $\vr_{c,p}$ are the radius and location of the planet, respectively.
As the planet's radius is significantly smaller than all other distances in the problem, the rectangular function could be naturally approximated by a Dirac distribution.
Furthermore, the planet inherently possess a zero vertical extension, which leads to next r.m.s scale height for the system $\left\{ \text{gas disc} + \text{planet}\right\}$: $H_{gp}=H_g/\sqrt{2}$ \footnote{We dropped the "sg" superscripts for clarity} (see Eq.~\ref{Eq: mean root square height}).
With all these observations, we can calculate the gravitational force density of the planet acting on the gas disc:
\begin{equation}
\begin{array}{lll}
\vec{f}_{2D}^{p\rightarrow g} &=& \displaystyle - \Sigma_g \iint\limits_{disc} \Sigma_p K_{gp} \, \vec{e}_s \, d^2 \vr'  \\
                              &=& \displaystyle - \Sigma_g  \frac{m_p}{s} \sqrt{\frac{2}{\pi}} \frac{1}{H_g} c^2 \exp\left(c^2\right) \left[ K_1(c^2) - K_0(c^2) \right] \vec{e}_{c,p}
\end{array}
\end{equation}
where $c^2= \frac{s^2}{4 H_g^2}$ and $\vec{e}_{c,p}=(\vr-\vr_{c,p})/||\vr-\vr_{c,p}||$ is the unit vector oriented from the planet to the fluid element located at $\vr$.
Above expression is exactly the force density of a planet acting on the gas disc \citep[Eq. 25]{muller_kley_2012}, which validates the consistency of our approach.
In other words, the planet-disc interaction kernel is nothing more than a special case of the Bessel prescription emphasised in this article.
This observation, which began as a curiosity, constitutes our first analytical test.

Another basic analytical test for assessing the correctness of our approach lies in the retrieve of the potentials associated to razor-thin discs, i.e when the scale height of the disc tends to zero, $H_g^{sg} \rightarrow 0$.
This is achieved simply by using the well known result,
\begin{equation}
\delta (z) = \lim_{H_g^{sg} \rightarrow 0} \frac{1}{\sqrt{2 \pi} H_g^{sg}} e^{-\frac{1}{2} \left(z/H_g^{sg}\right)^2}
\end{equation}
which, when used in Eq.~\ref{Eq: SG force kernel general}, permits to get the force kernel of a razor-thin disc:
\begin{equation}
\begin{array}{lll}
\K
      & = & \displaystyle \lim_{H_{g}^{sg} \rightarrow 0} \frac{1}{2 \pi} \frac{s}{H_g^{sg} H_g^{sg}} 
            \displaystyle  \, \, \iint\limits_{z, z'=-\infty}^{+\infty} 
            \displaystyle \frac{e^{-\frac{1}{2} \left(z/H_g^{sg}\right)^2} e^{-\frac{1}{2} \left(z'/H_g^{sg}\right)^2}}{\left(s^2+(z-z')^2\right)^{{3}/{2}}} \, dz \, dz'  \\
    & = & \displaystyle s \iint\limits_{z, z'=-\infty}^{+\infty} 
            \displaystyle \frac{\delta(z) \, \delta(z')}{\left(s^2+(z-z')^2\right)^{{3}/{2}}} \, dz \, dz'  \\ 
    & = & \displaystyle \frac{1}{|\vr-\vr'|^2}
\end{array}
\end{equation}
Such limit naturally applies to the kernel under its closed form (Eq.~\ref{Eq: SG force kernel exact}) by the use of the dominated convergence theorem, which permits to commute the vertical integrals and the limits.
This second simple analytical validation test shows that we expect to recover analytical and numerical results that apply to flat discs.
Consequently, in order to benchmark our Kernel and check our numerical implementation, we decided to compare in next Sects. the SG forces provided by our prescription against exact derivations for specific matter distributions and/or 3D numerical tests.
We primarily focus in next Sections on the accuracy of our spectral solver and the comparison between 3D simulations and different 2D gravity prescriptions, leaving for a future article the benchmark of the multi-fluid treatment.
We start by presenting our numerical setup and codes.

\subsection{Numerical method and codes}

The tests of this section were performed employing both the \FargoCPT \citep{2024_rometsch} and \NIRVANA \citep{2004_ziegler,2016_ziegler} codes. 

\FargoCPT is a 2D finite difference code with a staggered mesh, employing an advection scheme akin to finite volume methods. 
It solves the hydrodynamics equations using operator splitting and a second-order upwind scheme.
\FargoCPT and its variants, such as \textsc{FARGOCA}\xspace \citep{2014_lega}, \textsc{FARGOADSG}\xspace \citep{2008_baruteau}, and \textsc{FARGO3D}\xspace \citep{2016_llambay}, are built upon the Fargo code presented in \citep{2000_masset}.
It is formally accurate up to second-order in space and first-order in time and relies on artificial viscosity for shock smoothing \citep{2024_rometsch}. 
In this code the Bessel kernel was computed by means of full FFT methods, which requires to satisfy  $H/r=const.$ (See Sect.~\ref{sec: efficient numerical method}).
Thus, we used a logarithmic grid in the radial direction and a linear grid in the azimuthal direction.
We also made use of zero-padding for density in order to guarantee artificial periodicity in the radial direction.

The Bessel kernel was also implemented in the updated version of the \NIRVANA code, as presented in \citet{2020_gressel}.
Unlike our implementation in \FargoCPT, this time we used a semi-spectral method, which removes the constraint on linear scale heights, allowing for linear radial grids and greater generality  (See Sect.~\ref{sec: efficient numerical method}).
Finally, in Sect.~\ref{sec:gaussian cylinders test}, we also employed \NIRVANA to perform 3D simulations, enabling a comparison between the outcomes of 2D and 3D simulations. 
In particular, \NIRVANA features a new approach \citep{2024_gressel_ziegler} based on James' method for the computation of the gravitational potential at the poloidal domain boundaries \citep[see][]{1977_james,2019_moon}.
We start by showing that the Bessel prescription permits the retrieve the radial accelerations of flat power law discs.

\subsection{Power law discs}\label{sec: power law}

\begin{figure}[h]
\centering
\includegraphics[width=\hsize]{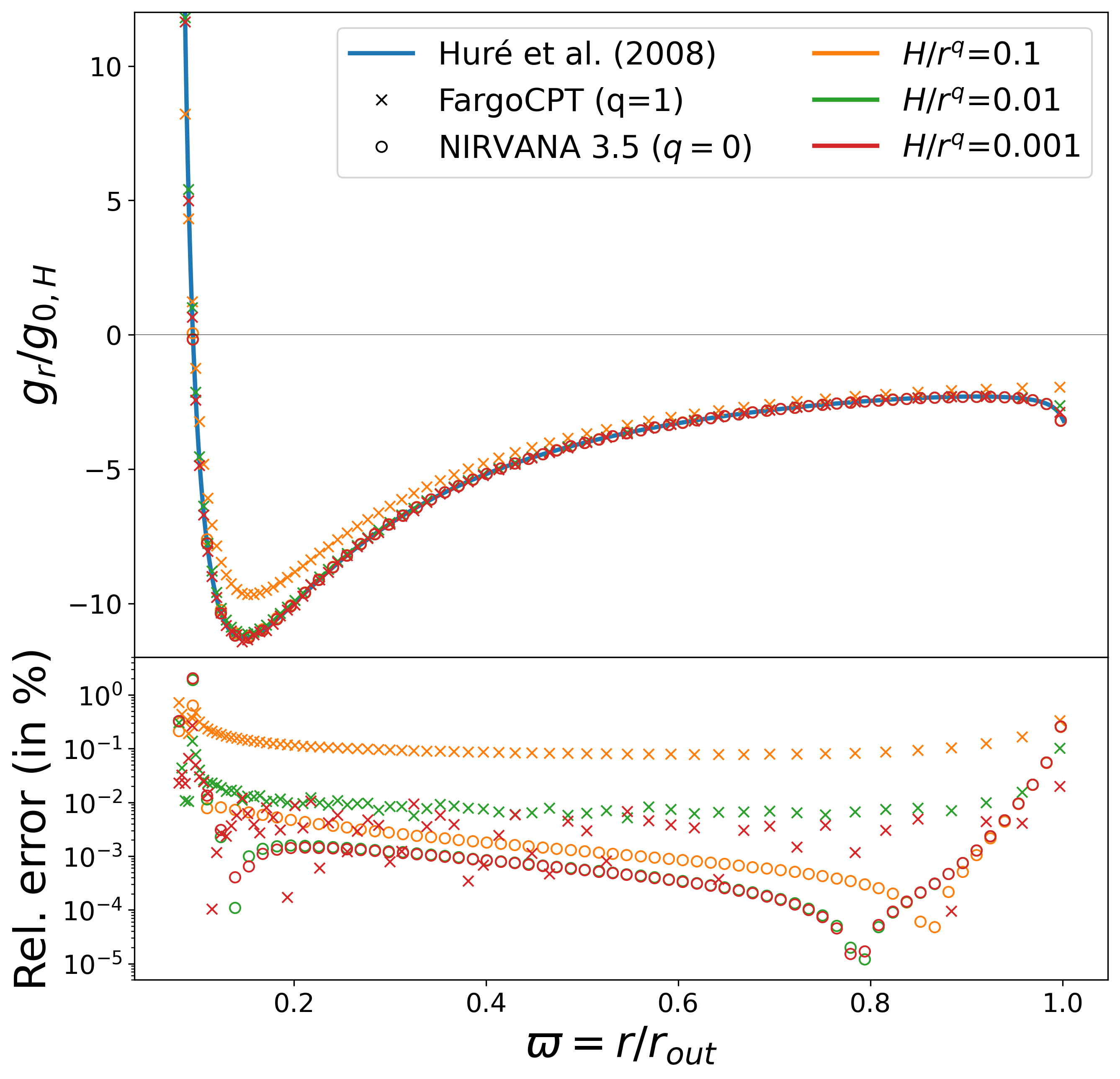}

\caption{Radial SG force of the power law disc for different scale heights. \\
The accuracy of our kernel increases with decreasing aspect ratios.
Note that for $\varpi\simeq0.07$ the relative error is beyond unity, which is a bias in the statistical measure due to a vanishing radial acceleration.
We recall that in \FargoCPT and \NIRVANA we used a logarithmic and uniform radial grid, respectively.
} 
\label{fig: power-law disc SG}
\end{figure}

One of the very few exact analytical solutions of the gravitational potential in flat discs, concerns axisymmetric, finite-size and power law density discs \citep{2007_hure_hersant, 2008_hure_hersant, baruteauPredictiveScenariosPlanetary2008a}.
For such a disc, the authors decomposed the exact potential in terms of series  which permits an evaluation of the SG radial component \footnote{This expression can be slightly different for specific values of the surface density power index $\beta_\rho$.}:
\begin{equation}
\displaystyle \frac{g_r^{(n)}(\varpi)}{g_{0,H}} = A \left( {1+\beta_\rho} \right) \varpi^{\beta_\rho} + \sum\limits_{k=0}^n \left( 2k a_k \varpi^{2 k - 1} - \left(2k+1\right) \frac{b_k}{\varpi^{2 (k+1)}} \right)
\end{equation}
where $\varpi=r/r_{out}$, $\beta_\rho$ is the surface density power index and $g_{0,H}=-2 \pi G \Sigma(r_{out})$.
The constant $A$ and the sequence coefficients can be found in the original papers.
Naturally, above series converges asymptotically to the exact gravitational force. 
In their study, the authors considered an infinitely thin disc and evaluated the potential in the equatorial plane.
Conversely, our results are applicable to a Gaussian vertical profile but they can be safely compared to theirs in the limit of the razor-thin disc approximation, i.e small aspect ratios: $H_g^{sg}/r \ll 1$.

In Fig.~\ref{fig: power-law disc SG}, we depicted the force derived from \citet{2008_hure_hersant} potential (blue solid line) for $n=100$ and compared it with the force provided by our Bessel kernel for the power law index $\beta_\rho=-1.5$ and decreasing aspect ratios.
Our numerical setup is:
\begin{equation}
\left\{
\begin{array}{lll}
\Sigma_0(r)        &=& 20000 \, (r/1\,\mbox{AU})^{\beta_\rho} \, \mbox{g.cm}^{-2}  \\
H/r^q              &=& [0.1, 0.01, 0.001] \\
(r_{in}, r_{out})  &=& (20, 250) \, \mbox{AU} \\
(N_r, N_\varphi)    &=& (1200, 2800)
\end{array}
\right.
\end{equation}
where $N_r$ and $N_\varphi$ are the number of cells in the radial and azimuthal direction, respectively.
Using FFT in the radial direction with \FargoCPT required setting $q=1$, while for \NIRVANA, which offers more flexibility, we used $q=0$.
We chose these two distinct values of $q$ to illustrate that our framework accommodates to any disc flaring.
For \FargoCPT simulations, the relative error decreases with a decreasing disc aspect ratio, aligning with our Bessel kernel's ability to achieve the flat disc limit. 
For smaller aspect ratios, the accuracy is limited to approximately $\sim 10^{-3}$ and $\sim 10^{-4}$ for \FargoCPT.
The same trend is observed with \NIRVANA, but the relative error reaches $10^{-4}$ more rapidly as the aspect ratio decreases.
Notably, when $H=0.1$, the error is around $10^{-3}$, two orders of magnitude smaller than the equivalent $H/r=0.1$ setup. 
This indicates that the common assumption of $H/r=const.$ in thin disc simulations may introduce significant errors.
We highlight that the exact radial SG force cancels at $\varpi\simeq0.07$, which leads to a statistical bias in the relative error in this region.
Finally, we checked that the accuracy trend with respect to the disc aspect ratio remains consistent for other power law indices ranging from $[-3,3]$, validating our first test.

\subsection{The exponential disc}\label{sec: exponential disc}

\begin{figure}[h]
\centering
\includegraphics[width=\hsize]{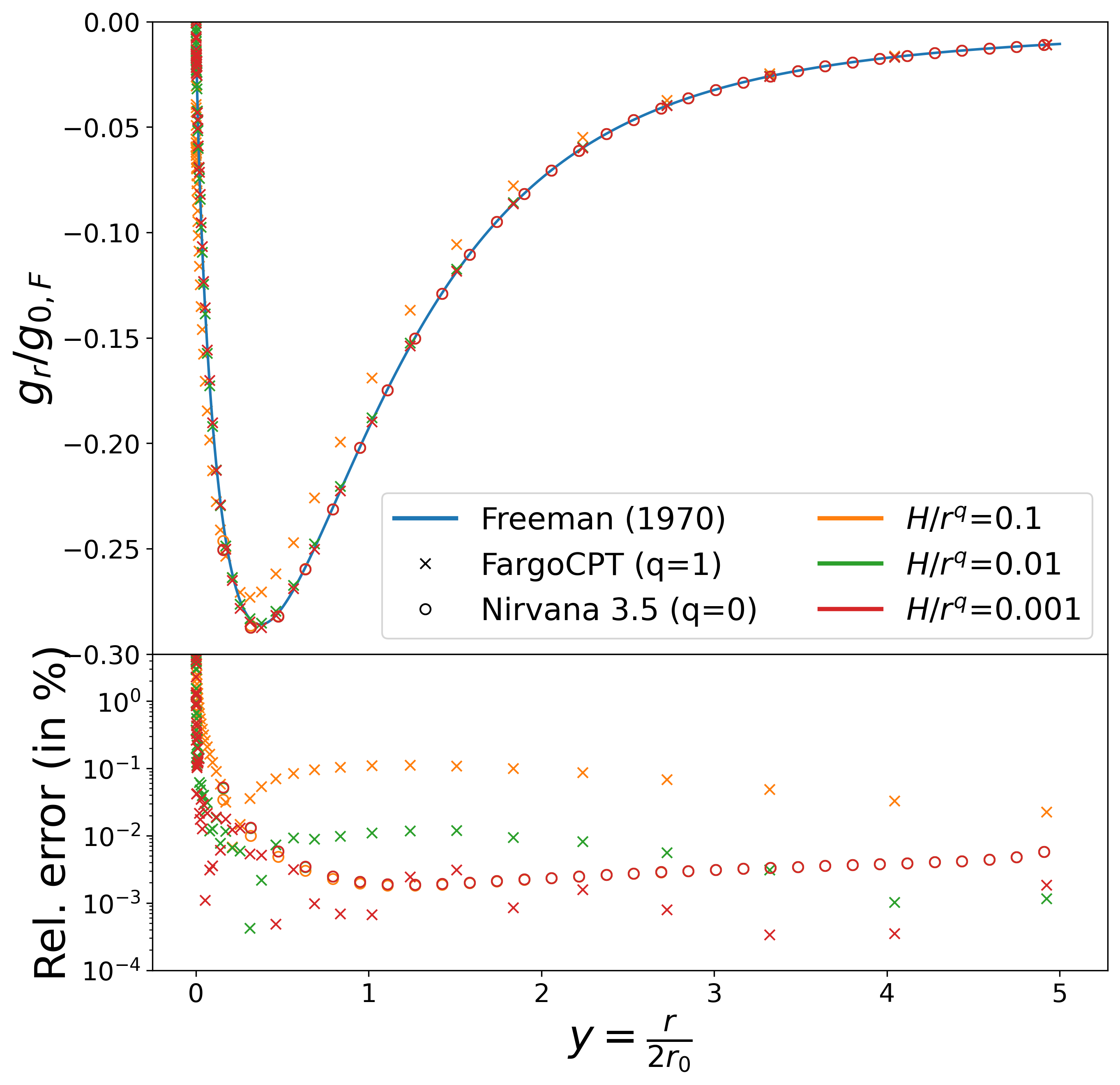}

\caption{Radial SG force of the exponential disc for different scale heights. 
The accuracy of our kernel increases with decreasing aspect ratios and piles up at $\sim 10^{-4}$. 
Note that for $y=0$ the relative error is beyond unity, which is a bias in the statistical measure due to a vanishing radial acceleration.
} 
\label{fig: exponential disc SG}
\end{figure}

Another analytical result that is useful for comparison with our Bessel kernel relates to exponential discs:
\begin{equation}
\Sigma(r) = \Sigma_0 \exp\left( -r/r_0 \right)
\end{equation}
Indeed, the gravitational potential associated to these discs in the equatorial plane is known \citep[Sect. 2.6]{1970_freeman, 2008_binney_tremaine}, which permits to derive exactly the radial acceleration due to the disc own gravity:
\begin{equation}\label{Eq: radial acceleration exponential disc}
g_r(r) = - g_{0,F} \, y \left[ I_0(y) K_0(y) - I_1(y) K_1(y) \right]
\end{equation}
where $y=r/(2 r_0)$ and $g_{0,F}= 2 \pi \, G \Sigma_0$.
The notations $I_\alpha$ and $K_\alpha$ refer to modified Bessel functions (of order $\alpha$) of the first and second kind, respectively.
Once again, our goal here is to compare the exact acceleration of the exponential discs (Eq.~\ref{Eq: radial acceleration exponential disc}) with the acceleration obtained numerically by the use of the Bessel prescription in the limit of small disc aspect ratios.
For the numerical side, our setup is:
\begin{equation}
\left\{
\begin{array}{lll}
\Sigma_0(r)        &=& 200  \, \mbox{g.cm}^{-2}  \\
H/r^q              &=& [0.1, 0.01, 0.001] \\
r_0                &=& 250 \, \mbox{AU} \\
(r_{in}, r_{out})  &=& (0.01, 2500) \, \mbox{AU} \\
(N_r, N_\varphi)    &=& (1200, 2800)
\end{array}
\right.
\end{equation}
We maintained the same setup as in the previous section, with $q=1$ for \FargoCPT and $q=0$ for \NIRVANA.
The results of this comparison are depicted in Fig.~\ref{fig: exponential disc SG}.
The conclusions are similar as the ones drew for the power law disc (see Sect.~\ref{sec: power law}), with the difference that this time the radial SG force vanishes at $y=0$.
The only difference is that \FargoCPT more closely approximates the analytical solution than \NIRVANA for $H/r^q=0.001$. 
This is because \FargoCPT's logarithmic grid better resolves the inner disc, where most of the mass lies.

\subsection{Gaussian discs test}\label{sec:gaussian cylinders test}

\begin{table}
\caption{Parameters for the Gaussian cylinder test} 
\label{tab: gaussian cylinder test}                 
\centering                                          
\begin{tabular}{c c c c}                            
\hline                                              
Amplitude ($A_i$) & Center location ($r_{c,i}, \varphi_{c,i}$) & Radius ($R_i$) \\ 
\hline                                               
1000    & (5.0,  $\pi$)      & 0.50   \\ 
200     & (4.5, $5\pi/3$)    & 0.50   \\ 
400     & (5.5,  $\pi/3$)    & 0.15   \\ 
\hline
\end{tabular}
\end{table}

\begin{table}
\caption{L$_\infty$ deviations used in the Gaussian disc test.} 
\label{tab: error gaussian discs test}            
\centering                                          
\begin{tabular}{l | c c c}                            
\hline                                              
                           & $H=0.4$           & $H=0.1$             & $H=0.01$            \\ 
\hline                                              
Bessel                     & $3\%$   & $3  \%$ & $5 \%$ \\ 
Plummer, $\epsilon/H=1.2$  & $28\%$  & $7  \%$ & $5 \%$ \\ 
Plummer, $\epsilon/H=0$    & $129\%$ & $32 \%$ & $8 \%$ \\ 
\hline
\end{tabular}
\end{table}

\begin{figure*}[!th!p]
\centering
\resizebox{0.95\hsize}{!}
{
    
    \includegraphics[width=0.9\hsize]{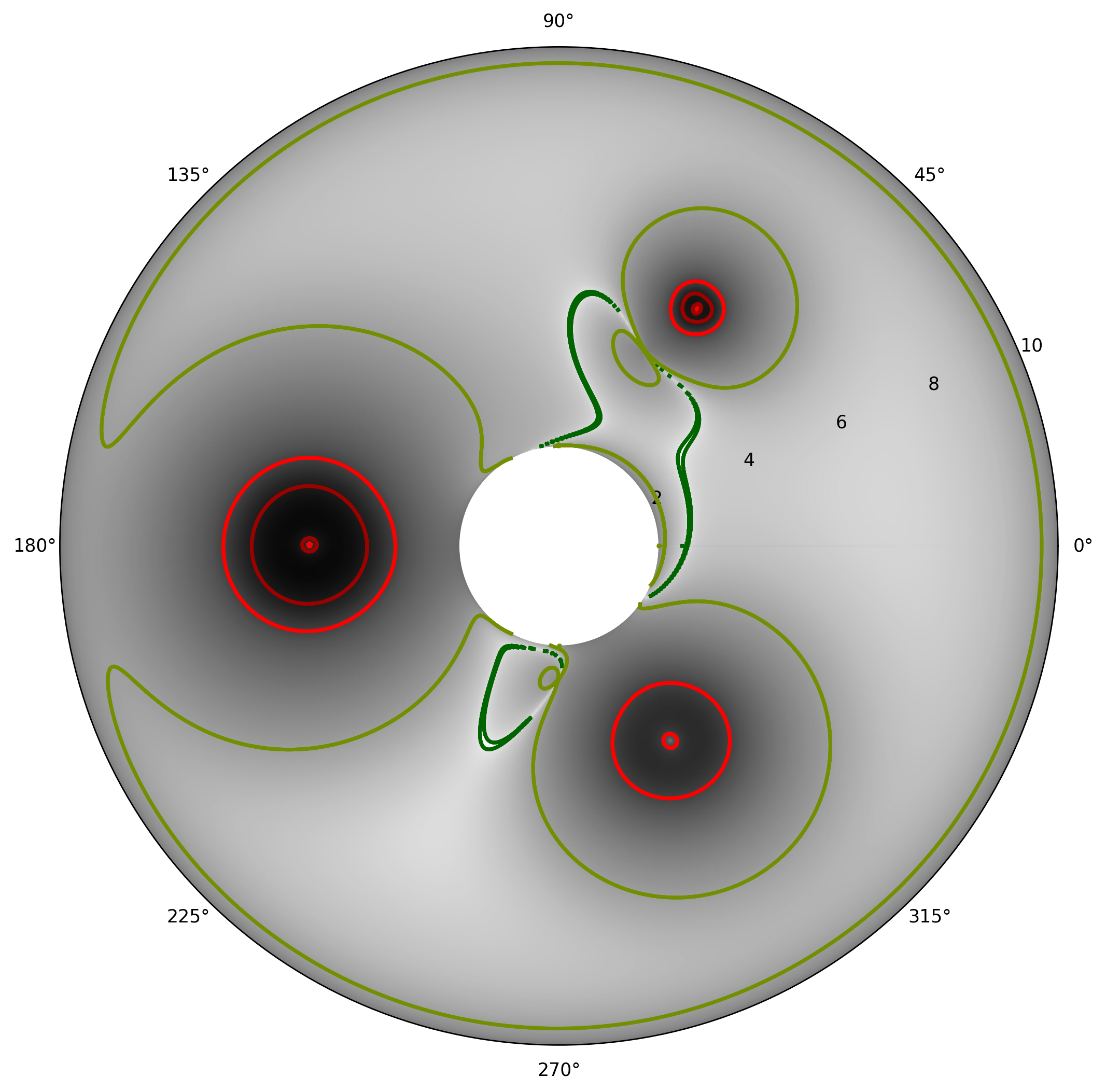}
    \includegraphics[width=0.9\hsize]{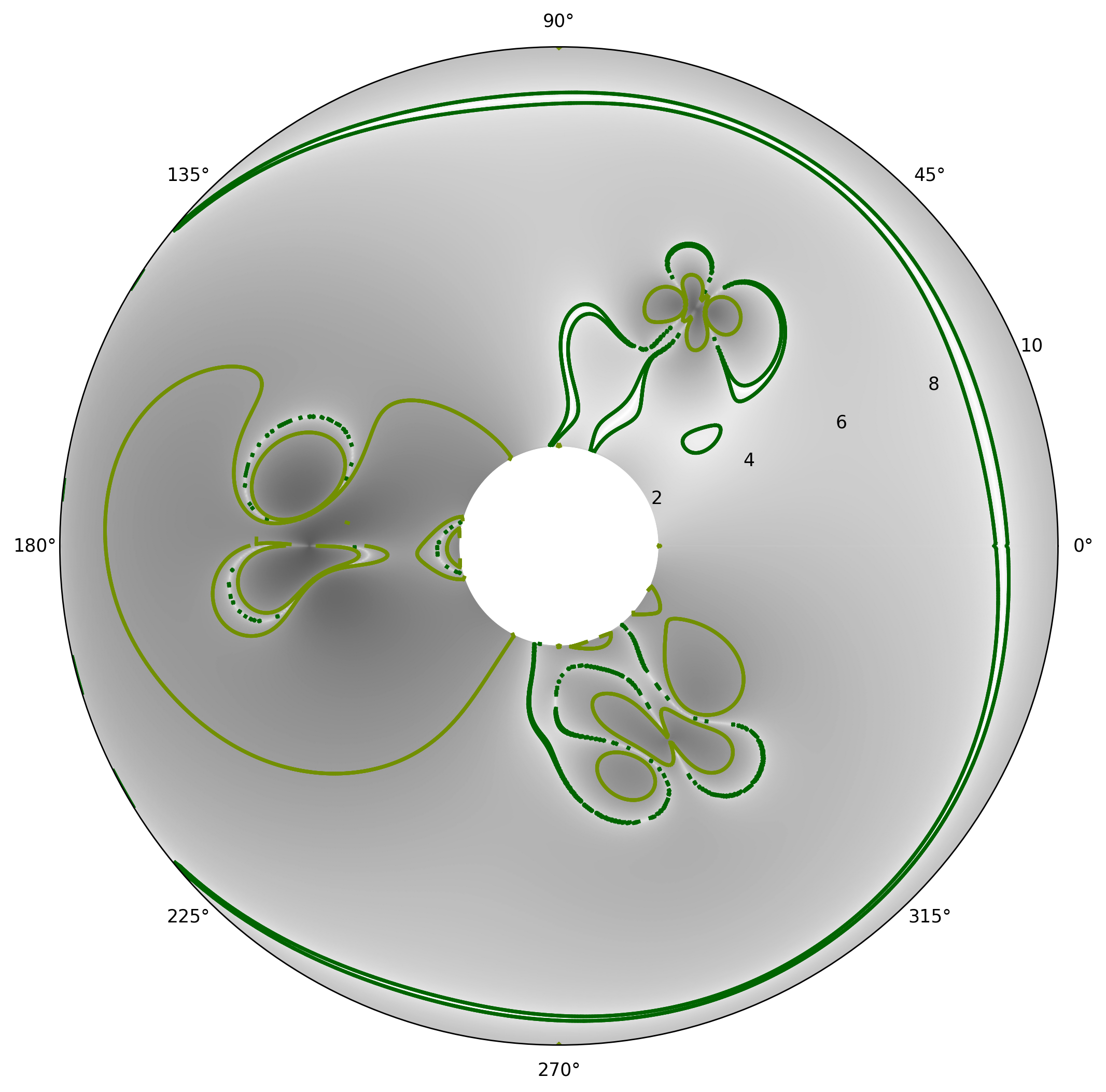}
    \includegraphics[width=0.9\hsize]{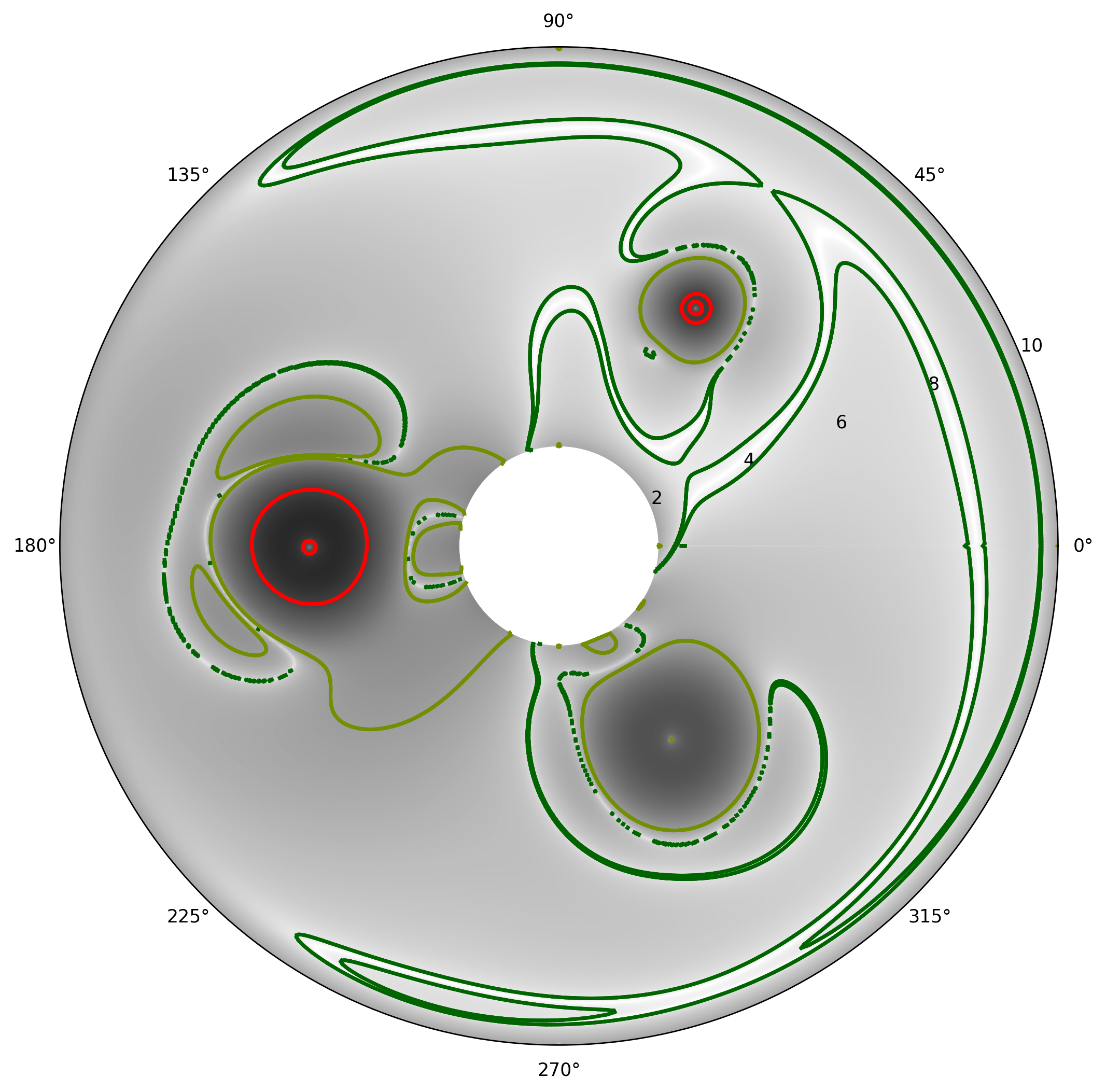}
}
\vfill
\resizebox{0.85\hsize}{!}
{
    \includegraphics[width=0.9\hsize]{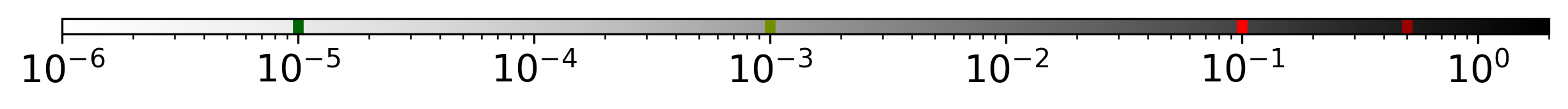}
}
\vfill
\caption{Relative difference between the 2D self-gravity prescriptions and the reference gravitational field (see Eq.~\ref{Eq: relative reference gaussian discs}) for $H=0.4$ AU.
From left to right: ~a) the sharp potential with $\epsilon/H=0$, ~b) the variable Bessel kernel, and ~c) the soft potential with $\epsilon/H=1.2$. 
The colours of the contour levels at $10^{-5}, 10^{-3}, 0.1, 0.5$ are indicated in the colorbar.
Softening prescriptions exhibit errors exceeding tens of percent at disc positions, whereas the Bessel prescription maintain errors below 3\%.
}
\label{fig: cylinders test}
\end{figure*}

To assess the accuracy of 2D SG prescriptions it is common to resort to the Gaussian spheres test \citep{chan_2006, shengtai_2009}.
This test involves averaging vertically the 3D gravitational field associated with 3D Gaussian spheres positioned at different locations and comparing it with the results obtained using a 2D kernel. 
A benefit of this benchmark is its capability to assess the azimuthal component of the gravitational force.
Given the analytical challenges in vertically averaging the SG force of a Gaussian sphere, we chose to conduct this test solely through numerical methods as explained in next paragraphs.

Our test entails computing the 3D gravitational field of Gaussian discs in $z={\rm const.}$ planes and serving as perturbation to a background Gaussian vertically stratified density.
The resulting 3D gravitational field is then vertically integrated and used for computing the reference field amplitude in the equatorial plane:
\begin{equation}\label{Eq: vertically integrated grav. field}
\mathcal{G}^{3D}_{\rm ref} = \sqrt{\Biggl{\langle} \frac{1}{\sqrt{2 \pi} H} e^{-\frac{1}{2}(z/H)^2} \mathcal{G}_r \Biggr{\rangle}_z^2 
+ \Biggl{\langle} \frac{1}{\sqrt{2 \pi} H} e^{-\frac{1}{2}(z/H)^2} \mathcal{G}_\varphi \Biggr{\rangle}_z^2}
\end{equation}
We highlight that $\mathcal{G}^{3D}_{\rm ref}$ denotes the 3D gravitational field obtained with \NIRVANA, followed by vertical integration denoted by $\langle\dots\rangle_z$, with the appropriate factors accounted for.
This reference field is then compared with the results of the 2D SG prescriptions, 
\begin{equation}
\mathcal{G}_{2D} = \left| \iint\limits_{disc} \, \Sigma_{g} \, \text{K}_{gg} \, \es \, d^2\vr' \right|,
\end{equation}
specifically the Bessel kernel, on the one hand, and a Plummer potential with two different SL values: $\epsilon/H=[0, 1.2]$, respectively, on the other hand.
While $\mathcal{G}_{2D}$ is not strictly the absolute value of a 2D gravitational field, as it is not derived from a potential (as detailed in Sect. \ref{subsec: flat disc poisson equation and gravitational potential}), it remains the quantity that can be faithfully compared with its 3D counterpart, $\mathcal{G}^{3D}_{\rm ref}$.
To assess the potential limitations of each prescription based on the disc vertical extent, we conducted this test for three different scale heights: $H=[0.01, 0.1, 0.4]$.

For this test, the density consists of a constant background distribution,
\begin{equation}
\left\{
\begin{array}{lll}
\rho_0(\vr, z) & = & \frac{\Sigma_0}{ \sqrt{2\pi} H} \exp\Big(-\frac{1}{2}\left(\frac{z}{H}\right)^2\Big) \\[4pt]
\Sigma_0       & = & 200 \, \mbox{g\,cm}^{-2} \\[4pt]
H              & = & [0.1, 0.01] \, \text{AU} \\
\end{array}
\right.
\end{equation}
with perturbations introduced in the form of aforementioned Gaussian discs,
\begin{equation}
\displaystyle \rho(\vr, z) = \rho_0(\vr,z) \prod\limits_{i=1}^{N=3}\left[ 1 + A_i \, \exp\left(- \frac{1}{2} \frac{|\vr-\vr_{c,i}|^2}{R_i^2} \right) \right]
\end{equation}
The parameters $A_i$, $r_{c,i}$ and $R_i$, that represent the relative amplitude, centre location and radius of the Gaussian, respectively, are detailed in Table~\ref{tab: gaussian cylinder test}.
For the evaluation of the 3D gravitational field employing the \NIRVANA code, we used a cylindrical grid with next computational domain extents:
\begin{equation}
\left\{
\begin{array}{lll}
r    & \in  & [2, 10] \, \text{AU}  \\
\varphi & \in  & [0, 2\pi] \\
z & \in & [-5 H, 5 H]
\end{array}
\right.
\end{equation}
We used two different number of cells for the 3D setup:
\begin{equation}
(N_r, N_\varphi, N_z)=
\left\{
\begin{array}{lll}
(256, 1024, 64) & \text{if}    & H=0.01 \, \text{AU} \\ [8pt]
(256, 1024, 128)  & \text{if}  & H=0.4, \, 0.1 \, \text{AU} 

\end{array}
\right.
\end{equation}
This distinct choice permits to reduce the anisotropy in our 3D grid, which facilitates the convergence of the multigrid solver.
For the 2D simulations counterpart, we kept the radial and azimuth resolution unchanged compared to the 3D setup.

In Fig.~\ref{fig: cylinders test} we depicted the relative difference between the reference simulation and it is 2D analogues for different SG prescriptions,
\begin{equation}\label{Eq: relative reference gaussian discs}
\varepsilon = \left|\mathcal{G}_{2D}-\mathcal{G}_{3D}^{ref}\right| / \max(\mathcal{G}_{3D}^{ref})
\end{equation}
associated to above setup and for $H=0.4$.
The amplitude of the 2D gravitational field, computed using the semi-spectral method implemented in \NIRVANA, is depicted by $\mathcal{G}_{2D}$.
The choice of above error stems from the fact that the gravitational field amplitude approaches 0 in two regions, which makes the standard relative norm diverge, which bias the interpretation.
We quantified the deviation by presenting following L$_\infty$ norm:
\begin{equation}
||\varepsilon||_\infty = \max\left(\varepsilon\right)
\end{equation}
Above $L_\infty$ norm is presented in Table~\ref{tab: error gaussian discs test} for various 2D SG prescriptions.
Our findings indicate that for the three different scale heights, the error associated with the Bessel kernel remains within a few percent.
Although the maximum error is found near the Gaussian discs positions, it is not centered on them.
These minimal deviations, even for high scale heights such as $H=0.4$, are due to our method's versatility, which is not limited to razor-thin discs but is applicable to any geometrical aspect ratio.
Conversely, the error associated with the Plummer potential varies with the disc scale height and decreases as the scale height diminishes.
Specifically, for $\epsilon/H=1.2$ and $H=0.4$, the maximum error reaches 28 \% and is located in the disc of the smallest radius.
Indeed, this prescription fails to resolve gravitationally bumps with dimensions smaller than the smoothing length. 
Notably, this setup generally underestimates SG.
On the other hand, for $\epsilon/H=0$ and $H=0.4$, the maximum error reaches 129 \%.
In this case the highest errors are located in the regions of highest density, and this prescription generally overestimates SG.
When the scale height decreases to $H=0.01$, all three prescriptions yield similar results, with an overall maximum error of just a few percent.

With this test, we have demonstrated that the Plummer potential prescription fails to provide an accurate estimation of SG in regions of high density, where SG is overestimated for $\epsilon/H=0$, and in regions smaller than the scale height for $\epsilon/H=1.2$, where SG is underestimated.
Conversely, the Bessel prescription consistently delivers accurate results across the entire range of scale heights.

\section{Discussion}\label{Sect: discussion}

In this section, we will explore the potential consequences of our approach for tackling GI, as well as the numerical methods for estimating SG, and we will consider alternative methods that offer greater generality.
Then, we identify the limitations of our approach and present solutions for scenarios where we deviate from the initial assumptions. 
Finally, we delve into a detailed discussion on the Poisson's equation for 2D systems and its implicit implications for 3D systems.
In this context, we propose an effective 2D gravitational potential consistent with our formalism.

\subsection{Consequences for planet formation}\label{sec: consequences planet formation}

Our detailed study of the Bessel kernel, along with the tests conducted in Sects.~\ref{sec: advantages associated with the Bessel kernel}-\ref{Sect: validation}, has revealed and quantified that the Plummer potential can both under- and overestimate the importance of SG.
These findings could have profound implications for planet formation theories, particularly the GI, as most simulations characterizing it have been conducted using a thin disc approximation. 
This approach typically employs a smoothing length prescription for the gravitational potential or involves solving a 2D Poisson equation.
We take this opportunity to clarify that solving a 2D Poisson equation, as often done in the flat disc limit, is equivalent to using a Plummer potential with $\epsilon/H=0$, as explained in Sect.~\ref{subsec: flat disc poisson equation and gravitational potential}.

On the one hand, we affirm that a finite smoothing length suppresses the Newtonian character of gravity, potentially quenching gravitational collapse or enhancing clump disruption during spiral encounters \citep{2015_young_clarke}.
On the other hand, we also affirm that a vanishing smoothing length artificially amplifies gravity.
In this context, we recall that an ongoing issue is the convergence problem, which stems from the lack of consensus about the threshold cooling parameter that separates the regime of gravito-turbulence from fragmentation in 2D simulations \citep{2015_young_clarke,2016_Kratter_Lodato}.
The issue first emerged in 3D SPH simulations of gravito-turbulence, where resolution-dependent artificial viscosity likely played a significant role \citep{2011_lodato_clarke}.
In 2D grid-based simulations, some aspects of the problem were later attributed to inner edge effects \citep{2011_paardekooper}, though the issue remains only partially addressed in these setups.
Consistent with these earlier findings, \citet{2012_paardekooper} demonstrated that fragmentation is a stochastic process, allowing discs to fragment at broad beta-cooling ranges and blurring the boundary between gravito-turbulence and fragmentation regimes.
Given that these simulations were conducted thanks to 2D shearing box simulations with a similar framework where SL equals zero, we think that our Bessel formulation would be key to addressing the convergence problem.
More specifically it will likely narrow and better constrain the beta cooling parameter range where stochastic fragmentation occurs.

\subsection{Numerical assessment with FFT methods}\label{sec: efficient numerical method}

The calculation of SG using the Bessel kernel remains compatible with FFT techniques. 
This compatibility arises from expressing the radial and azimuthal components of the SG force as convolution products \citep{2008_baruteau,2023_rendon_restrepo}. 
The explicit forms of the kernel components, ready for use in convolution operations, are provided in Appendix~\ref{app: convolution product}.
The main advantage of this method lies in the acceleration of the calculation since it requires $N_r N_\varphi \log{\left(N_r N_\varphi\right)}$ operations, in contrast with direct summation that requires $(N_r N_\varphi)^2$ operations.
This gain in performance requires nonetheless to obey some constraints:
First, in order to enable a description in the form of a convolution product, it is necessary that the radial grid is logarithmic.
This requires that the disc aspect ratio of the self-gravitating disc, $H_{ab}^{sg}/r$, remains constant in the whole numerical domain (see appendix \ref{app: limitations of the full spectral method} for a discussion when deviating from this condition).
Following the procedure of \citet[Sect. 4.2]{2023_rendon_restrepo}, we found that this indirectly implies that the bi-fluid Toomre's parameter, $\TQ$, and the standard disc scale height, $H_a/r$, should remain constant throughout the entire numerical domain-- and also in time.
This stringent requirement precludes the use of FFT methods in realistic setups. 
Consequently, if FFT methods are necessary, the simplest approach is to use the thermal scale height without the SG correction: $H_{sg} \simeq H$. 
Second, the use of Fourier transforms demands periodicity in all directions, which requires to double the radial domain with a zero-padding in order to perform an aperiodic convolution \citep{hockney2021computer}.
This operation does not affect the computational endeavour since the Hermitian property of Fourier transforms halves the total number of transforms and, as a consequence, compensates the doubling of the domain.
Finally, this zero-padding does not affect the SG computation as checked in Sect.~\ref{Sect: validation}.

Computing the Bessel kernel requires using special functions which is computationally expensive.
In \FargoCPT, the Bessel kernel calculation is roughly 20 times slower than the original kernel calculation.
If the Bessel kernel is recomputed at every iteration, this increases the cost of SG from around 40\% to around 130\% of the transport step cost, or from 21\% to 45\% of the total computational costs when using a locally isothermal equation of state.
For the latter, however, the SG kernel can be precomputed for the whole simulation because it does not change with time.
Then, there is no difference in computational cost after the initialization.
In case of a non-isothermal equation of state, the kernel can be recomputed only every so often to save computational costs. This can be justified by the fact that the scale height of the disk only changes slowly, at least when the cooling timescale is short compared to the dynamical timescale.
For a more realistic stratification in presence of SG, there is however also a physical issue. 
In this case, the aspect-ratio of the disk is no longer constant and the method is formally no longer applicable and only yields an approximation to the SG accelerations.

In general, we expect the disc flaring not to be constant, which renders the application of our method with FFT impractical.
For instance, the $\Hab^{sg}/r={\rm const.}$ constraint (necessary for full spectral methods) becomes prohibitive when investigating massive discs. 
Indeed this constraint implies that the Toomre's parameter needs to be constant in the whole domain and frozen in time.
Therefore, a potential decrease in the Toomre's parameter due to a gravitational runaway process will not be captured by these methods.
As a consequence, the other computational possibility, that was in fact implemented in \NIRVANA, consists on resorting to a pseudo-spectral method based on a combination of FFT in the azimuthal direction followed by a direct summation in radial direction, whose complexity is in $N_r N_\varphi\log\left( N_\varphi \right) + N_\varphi N_r^2$ \citep{shengtai_2009}.
What is new in this implementation is that every few time steps the scale height is updated according to Eq.~\ref{Eq: stratification parameters}.
This method is a good trade-off between computational efficiency and generality, since it is adapted to any scale height prescription beyond problems governed only by self-gravity.
Another possible solution that eliminates the $\Hab(\vr)/r={\rm const.}$ limitation while preserving the computational efficiency, with $N \log{N}$ operations, involves using a Fast Hankel transform algorithm in the radial direction.

We will now address the numerical treatment of the multi-fluid case.
In this scenario, it is necessary to pre-compute the self-kernels for each fluid, as well as the cross-gravitational terms between different fluid species.
This can be quite burdensome.
However, since the Kernels ultimately depend only on the root mean square scale height, we propose that some of the involved fluids are assumed to have the same scale heights.
This significantly reduces the number of required Kernels.
For instance, consider a situation where two dust species have the same scale height. In this case, only four Kernels and six convolutions would need to be computed at each time step:

\noindent 
\textbf{Volume forces acting on the gas:}
\begin{equation}
\Sigma_g \, \fourier^{-1}\left[\hat{\Sigma}_g \ast
\hat{\text{K}}_{gg}\right] 
\quad \text{and} \quad 
\Sigma_g \, \fourier^{-1}\left[(\hat{\Sigma}_{d1} + \hat{\Sigma}_{d2}) \ast \hat{\text{K}}_{dg}\right] 
,\end{equation}
\textbf{Forces acting on the dust 1:}
\begin{equation}
\Sigma_{d1} \, \fourier^{-1}\left[(\hat{\Sigma}_{d1} + \hat{\Sigma}_{d2}) \ast
\hat{\text{K}}_{dd} \right]
\quad \mbox{and} \quad
\Sigma_{d1} \, \fourier^{-1}\left[ \hat{\Sigma}_g \ast
\hat{\text{K}}_{gd} \right]
,\end{equation}
\textbf{Forces acting on the dust 2:}
\begin{equation}
\Sigma_{d2} \, \fourier^{-1}\left[(\hat{\Sigma}_{d1} + \hat{\Sigma}_{d2}) \ast
\hat{\text{K}}_{dd} \right]
\quad \mbox{and} \quad
\Sigma_{d2} \, \fourier^{-1}\left[ \hat{\Sigma}_g \ast
\hat{\text{K}}_{gd} \right]
.\end{equation}
Here, $\hat{}$ denotes the Fourier transform, $\fourier^{-1}$ the inverse Fourier transform, and $\ast$ the convolution product.
Accuracy of the approximation depends on the chosen scale height for calculating the dust kernels. A wise choice would be to take the harmonic mean of the involved fluids. In the above example, this would be: $H_d=\left( \frac{1}{H_{d1}} + \frac{1}{H_{d2}} \right)^{-1}$.

\subsection{Limitations}

The use of the Bessel kernel proposed in this paper requires that the stratification of the disc is strictly Gaussian.
This is of course an assumption that depends on the physical ingredients incorporated in the simulated disc and can be altered, for example, by considering the stellar irradiation \citep{2021_wu_lithwick} or in the presence of magnetic fields \citep[e.g.,][]{2008_johansen,2012_gaburov}.
In the former case, the system is constantly evolving in the vertical direction which suppresses a potential vertical hydrostatic equilibrium.
However the Bessel kernel does not require hydrostatic equilibrium. 
It only requires a Gaussian stratification, which is possible even when the equilibrium is not ensured in the vertical direction.
Once again all the time evolution information should be incorporated in the scale height.
For the latter, a simple rectification would be to replace the sound speed of gas by $c_g (1+1/\beta)^{1/2}$, where $\beta$ is the plasma beta parameter \citep{2008_johansen}. This would only modify the standard scale height $H_g$ used in Eq.~\ref{Eq: stratification parameters}.
In the case of more sophisticated stratifications that significantly deviate from the Gaussian one, a reasonable approach to maintain the applicability of our prescription would be to decompose the vertical profile of the disc as a finite linear combination of Gaussian functions. 
It is important to note that this method will result in an increasing number of numerical evaluations by FFT methods. 
If such a case occurs, it is surely wiser to conduct 3D simulations directly.

During our numerical tests, we observed the generation of NaN values in the SG forces when dealing with small disc aspect ratios.
This occurrence was traced back to the overflow of the exponential function employed in the Bessel kernel definition. 
This issue is simply addressed by using the Taylor expansion of the kernel at infinity:
\begin{equation}
\K(X) = \frac{1}{2\!\sqrt{2}} \frac{\Hab^{-2}}{d_{ab}} 
\left[ \frac{1}{X^{1/2}} -\frac{3}{8} \frac{1}{X^{3/2}} + \frac{45}{128} \frac{1}{X^{5/2}} + \mathcal{O}\left(\frac{1}{X^{7/2}} \right) \right]
\end{equation}
where $X=d_{ab}^2/8$.
We applied above approximation when the variable $X$ was larger than 50, successfully resolving our issue without any impact on the accuracy of the method.

\subsection{Indirect term}

The presence of asymmetries in the disc can lead to an offset between the barycenter of mass and the reference frame, located at the star's position.
This offset is accounted through an indirect term: $\Phi_{ind}(\vr)=- \vec{a}_{\star}\cdot \vec{r}$, where $\Vec{a}_{\star}$ is the acceleration of the central object \citep{2016_zhu, regaly_vorobyov_2017, rendon_2022}.
In order to be compatible with this investigation, we found relevant to derive the indirect term associated to the pull exerted by fluid `a' on the star using our approach:
\begin{equation}\label{Eq: indirect term}
\Phi_{a, \text{ind}} (\vr) = \sqrt{2} r \iint\limits_{disc} \Sigma_a(\vr')  \cos{\left(\varphi-\varphi'\right)} \text{K}_a(d_a) d^2 \vr' 
\end{equation}
where $\displaystyle \text{K}_a(d_a)=\frac{1}{\sqrt{\pi}} (H_a^{sg})^{-2} \frac{d_a}{4} \exp\left(\frac{d_a^2}{4} \right) \left[ K_1\left(\frac{d_a^2}{4} \right) - K_0\left(\frac{d_a^2}{4} \right) \right]$ and $d_a=||\vr'||/H_a^{sg}$.
In appendix~\ref{app: indirect term} we propose a detailed derivation where you can also find the gradient of this potential.
This term is particularly important since it was recently pointed out that the indirect term and the SG forces are inseparable ingredients that should be accounted, or disregarded, simultaneously in simulations \citep{2025_crida}.

\subsection{Flat disc Poisson equation and gravitational potential}\label{subsec: flat disc poisson equation and gravitational potential}

For numerical simulations addressing SG in thin discs, it is common to solve next Poisson's equation \citep{2001_gammie},
\begin{equation}\label{Eq: Poisson eq. dirac}
\Delta \Phi = 4 \pi G \Sigma \delta(z)
\end{equation}
or to assess an integral \citep[cf.][]{2012_paardekooper},
\begin{equation}\label{Eq: Potential integral dirac}
\Phi(x,y,z=0) = -G \int \frac{\Sigma \, \delta(z')}{\sqrt{(x-x')^2+(y-y')^2+z'^2}} dx' \, dy' \, dz'
\end{equation}
in order to evaluate the gravitational potential, $\Phi$.
It is worth reminding that solving above equations places us automatically in a paradigm where the disc is considered as infinitely thin, suggesting that the disc is cold and/or the disc SG dominate in the vertical direction.
Therefore, when above equation is used in 2D simulations, there is a possibility of overestimating the impact of SG, potentially resulting in inconsistent setups.
For instance, in the context of Gravitational Instability (GI), Toomre's parameters are often in the range of $0.5-1$.
This may pose a contradiction with the razor-thin disc approximation implied by Eqs.~\ref{Eq: Poisson eq. dirac}-\ref{Eq: Potential integral dirac}, which rather implicitly suggests $Q \ll 1$.
Further, there is several evidence that protoplanetary discs are thin but far from being flat, even in the dust phase, as supported by observations \citep{2023_lin}.
These two statements encourage us to think that the method presented in this paper is appropriate.
Indeed, in contrast to Eqs.~\ref{Eq: Poisson eq. dirac}-\ref{Eq: Potential integral dirac}, which are limited to razor-thin discs, the Bessel prescription is applicable to any Gaussian stratification of thin discs, ranging from razor-thin to thick discs, in alignment with observations.

In this context, it is legitimate to verify the possibility of existence of a Poisson's equation with our formalism.
Therefore, it is useful to rewrite the equivalent gravitational field in 2D:
\begin{equation}\label{Eq: grav. field 2D general}
\begin{array}{lll}
\Vec{\mathcal{G}}_{2D}^{a\rightarrow b} &=& \displaystyle - \int\limits_{z=-\infty}^{\infty} \frac{e^{-\frac{1}{2}(z/H_b^{sg})^2}}{\sqrt{2 \pi} H_{b}^{sg}} \Vec{\Nabla}_{3D} \Phi_a(\vr,z) \, dz \\
                       &=& \displaystyle - \int\limits_{z=-\infty}^{\infty} \frac{e^{-\frac{1}{2}(z/H_b^{sg})^2}}{\sqrt{2 \pi} H_{b}^{sg}} \Vec{\Nabla}_{2D} \Phi_a(\vr,z) \, dz
\end{array}
\end{equation}
where we used the even symmetry of the volume density with respect to the $z=0$
plane. 
In above expression, $\Phi_a$ is the potential generated by the mass distribution of the unique fluid `a'.
Therefore, the total 2D SG force per unit volume exerted by a disc made of two fluids `a' and `b' on a volume element of fluid `b' is: 
\begin{equation}\label{Eq: SG force general rewriting}
\begin{array}{lll}
\vec{f}^{tot \rightarrow b}_{2D}(\vr) = \Sigma_b \left( \Vec{\mathcal{G}}_{2D}^{a\rightarrow b} + \Vec{\mathcal{G}}_{2D}^{b\rightarrow b} \right) = \Sigma_b \Vec{\mathcal{G}}_{2D}^{tot, b}
\end{array}
\end{equation}
This expression, is another way of writing Eqs.\ref{Eq:3D volume force total}-\ref{Eq: SG force general}.
In general the 2D Nabla operator and the integral cannot commute since the scale heights $H_a^{sg}$ and $H_b^{sg}$ are $\vr$-dependent functions.
As a consequence, it is immediate that in our framework the SG force is not conservative since $\curl{\Vec{\mathcal{G}}_{2D}^{tot\rightarrow b}} \neq \vec{0}$.
This is not immediately dramatic -- or in contradiction with the underlying physics -- as there is no theoretical basis suggesting that the weighted average of a conservative force must retain its conservative nature. 
It would be tempting to consider that a force derived from a Plummer potential would better fit the Physics since it is conservative.
However, we remind that prior computations of SG with a softening prescription don't respect this condition since the gravitational acceleration is not a gradient of the Plummer potential.
Specifically, this is due to the absence of a $\Vec{\nabla} \epsilon^2$, where $\epsilon$ is the SL, term in the definition of the acceleration. 
This issue, already highlighted by \citet[Sect. 2.2]{2023_rendon_restrepo}, encourages to think that it is not necessary to use a conservative 2D force for fitting the vertical integral of a conservative 3D force.

Nonetheless, there is one scenario in which the vertically averaged SG force is conservative, and that is when $H_a^{sg}$ and $H_b^{sg}$ are spatial constants.
In such case, the gravitational field can be expressed in a conservative form:
\begin{equation}
\begin{array}{lll}
\Vec{\mathcal{G}}_{2D}^{a\rightarrow b} &=& - \Vec{\Nabla}_{2D} \Phi_{2D}^{ab} (\vr)                                        
\end{array}
\end{equation}
where the combined potential can be written in a closed form:
\begin{equation}\label{Eq: grav. potential 2D}
\begin{array}{lll}
\Phi_{2D}^{ab}(\vr) & = &  - \displaystyle \iint\limits_{disc}  \frac{\Sigma_a(\vr')}{2 \sqrt{\pi} \Hab} \exp\left( \frac{\dab^2}{8} \right) K_0 \left(\frac{\dab^2}{8} \right) d^2\vr' 
\end{array}
\end{equation}
The detailed derivation can be found in appendix~\ref{app: conservative force}.
We checked that the gradient of above potential permits to retrieve the SG force obtained in Eq.~\ref{Eq: volume force 2D}, which is consistent.
The right hand side integrand of Eq.~\ref{Eq: grav. potential 2D} can be interpreted as a Green's function, which scales as a logarithm or inverse law at the singularity or infinity, respectively.
This outcome is in agreement with the observations pointed out in Sect.~\ref{sec: nature of gravity ?}, which suggests that this Bessel potential permits a smooth transition for gravity between 2D and 3D.
We tested several combinations of differential linear operators to the potential defined by Eq.~\ref{Eq: grav. potential 2D}, in order to find a Poisson's like equation.
This task was unsuccessful but we still think that it is possible provided that more sophisticated mathematics tools are explored, like elliptic operators.
This is beyond the scope of our investigation.
While the constraint of a constant scale height is unrealistic \citep{2023_paneque_carreno} and limits the practical use of Eq. \ref{Eq: grav. potential 2D} in global simulations, it could still be valuable for 2D shearing box simulations. 
Especially if its Fourier transform is determined.
Finally, we want to highlight that in practice, simulations involving two distinct fluids with different scale heights, $H_g \neq H_d$,  will require to estimate four potentials: $\Phi_{2D}^{gg}$, $\Phi_{2D}^{dd}$, $\Phi_{2D}^{gd}$ and $\Phi_{2D}^{dg}$.

\section{Conclusion}

In this work, we proposed an exact self-gravity kernel readily to be used in 2D numerical simulations of PPDs made of gas and dust.
The closed form of this Bessel kernel results from a direct vertical double integration and its expression is:
\begin{equation}
\begin{array}{lll}
\K & = & \displaystyle \frac{1}{\sqrt{\pi}}\left( \Hab^{sg} \right)^{-2} \frac{\dab}{8} 
         \displaystyle \exp\left(\frac{\dab^2}{8} \right) \left[ K_1\left(\frac{\dab^2}{8} \right) - K_0\left(\frac{\dab^2}{8} \right) \right]
\end{array}
\end{equation}
This prescription applies when the vertical stratification of both fluids is Gaussian and all information regarding the vertical hydrostatic equilibrium due to the star's gravity and the disentangled gravity interaction of both fluids is encapsulated in the definition of the different scale heights $\Hab^{sg}$, $H_a^{sg}$ and $H_b^{sg}$ (Eqs.~\ref{Eq: stratification parameters}-\ref{Eq: mean root square height}).
In contrast with other kernels, the dependence of these scale heights with respect to the generalised Toomre's parameter, $\TQ$ (Eq.~\ref{Eq: stratification parameters}), allows a continuous transition from a regime of weak SG to a regime dominated by SG.
Additionally, this kernel captures correctly the short-range interaction of SG permitting a possible gravitational runaway and faithful 2D simulations incorporating SG.
Further, this kernel is intrinsically symmetric with respect to $\vr-\vr'$, which adheres to Newton's third law.
At our knowledge, this is the only 2D kernel in the literature of PPDs that simultaneously respects above three conditions.

On numerical grounds, the Bessel kernel is compatible with FFT methods provided that $\Hab^{sg}/r$ is constant over the computational domain.
A constraint that we lifted with a semi-spectral method in our \NIRVANA implementation.
We tested the analytical consistency and numerical implementation of this kernel by retrieving radial accelerations from the literature of flat discs, namely power law and exponential discs.
The relative errors are of the order of $10^{-4}-10^{-3}$, which also validates the accuracy of our FFT method despite the required zero-padding.
Furthermore, we compared the outcomes of 3D simulations with those of 2D simulations. This comparison highlighted the accuracy of our Bessel prescription and demonstrated that the softening prescription either underestimates or overestimates SG, with errors reaching up to 129 \%.

In agreement with earlier findings, our results indicate that for short distances the Plummer potential can lead to a significant underestimation of SG.
Conversely, a Plummer potential with smoothing lengths approaching zero, or equivalently solving a 2D Poisson equation, overestimates SG effects.
Both of these behaviours stem from the nature of SG within the Plummer potential framework, where the SL defines the scale below which gravity is screened.
In its absence, gravity retains its purely 3D form (scaling as $1/r^2$), which is not always physically appropriate in a 2D context.
In the Bessel prescription, a characteristic length also appears, but it represents the distance at which the gravitational force transitions—while retaining its Newtonian character—from a 3D scaling to a 2D scaling $(1/r)$.
As a consequence, we assert that the Bessel kernel proposed in this work should be used in order to conduct fully consistent 2D simulations including SG.
The SL approach, or 2D Poisson equation, widely used in PPDs studies may significantly misestimate SG, necessitating a reevaluation of these studies.
We will provide evidence for this statement in the accompanying article, where we will investigate the implications of our kernel to the GI paradigm of planet formation thanks to 2D global simulations.

\begin{acknowledgements}
We thank the referee, Clément Baruteau, for his insightful and constructive comments.
Funded by the European Union (ERC, Epoch-of-Taurus, 101043302). Views and opinions expressed are however those of the author(s) only and do not necessarily reflect those of the European Union or the European Research Council. Neither the European Union nor the granting authority can be held responsible for them. 
\end{acknowledgements}

\bibliographystyle{aa}
\bibliography{bibliography}

\begin{appendix}

\section{An useful function}\label{app:useful function}

Let's define the function sequence:
\begin{equation}\label{Eq: function Fn}
F_{n}(x)= \displaystyle \int\limits_{-\infty}^{+\infty}
        \frac{e^{-\frac{\xi^2}{2}}}{\left(x+\xi^2\right)^{n+\frac{1}{2}}} d\xi
\end{equation}
for $n \in \mathbb{N}$ and $x \in \mathbb{R}^{*}_+$.
This function satisfies following recursive relation:
\begin{equation}\label{Eq: recursive derivative}
\begin{array}{lll}
F_{n}(x) &=& -\frac{2}{2n-1} F_{n-1}^{(1)}(x) \\ [6pt]
         &=& \left(-\frac{2}{2n-1}\right) \left(-\frac{2}{2n-3}\right) ... \left(-\frac{2}{3}\right) \left(-\frac{2}{1}\right) F_{0}^{(n)}(x) \\  [6pt]
         &=& \displaystyle (-1)^n \frac{2^{2n}}{(2n)!} \, n! \, F_0^{(n)}(x)
\end{array}
\end{equation}
But:
\begin{equation}\label{Eq: function n=0}
\begin{array}{lll}
F_0(x) &=& \displaystyle \int\limits_{-\infty}^{+\infty}
           \frac{e^{-\frac{\xi^2}{2}}}{\left(x+\xi^2\right)^{\frac{1}{2}}} d\xi \\ 
       &=& \displaystyle \int\limits_{-\infty}^{+\infty} 
           e^{-\frac{x}{2} \sinh^2{t}} dt   \qquad \mbox{with} \quad \xi=\sqrt{x} \sinh{t} \\
       &=& \displaystyle e^{\frac{x}{4}} K_0\left(\frac{x}{4}\right)
\end{array}
\end{equation}
where $K_0$ is a modified Bessel function of the second kind and order 0.
Combining Eqs~\ref{Eq: recursive derivative} and~\ref{Eq: function n=0}, we get:
\begin{equation}\label{Eq: F3 definition}
\begin{array}{lll}
F_1(x) &=& \displaystyle (-1)^1 \frac{2^2}{2!} 1! \frac{d F_0(x)}{dx} \\ [8pt]
       &=& \displaystyle \frac{1}{2} \displaystyle e^\frac{x}{4} \left[ K_1\left(\frac{x}{4}\right) - K_0\left(\frac{x}{4}\right) \right] 
\end{array}
\end{equation}
where $K_1$ is a modified Bessel function of the second kind and order 1.

\section{Analytic expression of the self-gravity kernel}\label{app: integration}

In order to find a close form for the SG kernel, it is useful to compute the following quantity:
\begin{equation}
\begin{array}{lll}
I_{ab} &=&\displaystyle  \, \, \iint\limits_{z, z'=-\infty}^{+\infty} 
            \displaystyle \frac{e^{-\frac{1}{2} z^2/H_b^2} e^{-\frac{1}{2} z'^2/H_a^2}}{\left(s^2+(z-z')^2\right)^{{3}/{2}}} \, dz \, dz'  
            \\
\end{array}
\end{equation}
In this whole appendix Sect. we skip the "sg" superscripts for clarity.
Using a variable substitution we get:
\begin{equation}
I_{ab} =\displaystyle  H_a H_b \, \iint\limits_{u, v=-\infty}^{+\infty} 
            \displaystyle \frac{e^{-\frac{1}{2} u^2} e^{-\frac{1}{2} v^2}}{\left(s^2+(H_b u-H_a v)^2\right)^{{3}/{2}}} \, du \, dv  
\end{equation}
Above integral can be evaluated using following variable substitution \citep{shengtai_2009},
\begin{equation}
\begin{array}{cc}
\displaystyle 
\xi = \frac{H_b u - H_a v}{\sqrt{H_a^2+H_b^2}}, &
\displaystyle 
\xi' = \frac{H_a u + H_b v}{\sqrt{H_a^2+H_b^2}}
\end{array}
\end{equation}
which permits to get:
\begin{equation}
\begin{array}{lll}
I_{ab} 
   &=& \displaystyle \frac{H_a H_b}{\left( H_a^2 + H_b^2 \right)^\frac{3}{2}} 
       \displaystyle \iint\limits_{\xi, \xi'=-\infty}^{+\infty}
       \frac{e^{-\frac{\xi^2}{2}} e^{-\frac{\xi'^2}{2}}}{\left(\dsqbytwo+\xi^2\right)^{{3}/{2}}} d\xi d\xi' \\
   &=& \displaystyle \sqrt{2 \pi} \frac{H_a H_b}{\left( H_a^2 + H_b^2 \right)^\frac{3}{2}} 
       \displaystyle \int\limits_{\xi=-\infty}^{+\infty}
       \frac{e^{-\frac{\xi^2}{2}}}{\left(\dsqbytwo+\xi^2\right)^{{3}/{2}}} d\xi \\ [20pt]
   &=&  \displaystyle \frac{\sqrt{\pi}}{2} \frac{H_a H_b}{\Hab^3} F_1\left(\dsqbytwo\right)  \\
\end{array}
\end{equation}
where the function $F_1$ is defined by Eq.~\ref{Eq: function Fn} and it's closed form given by Eq.~\ref{Eq: F3 definition}.
Finally, combining above result with Eq.~\ref{Eq: SG force kernel general}, we get:
\begin{equation}
\K =  \displaystyle \frac{1}{\sqrt{\pi}}\left( \Hab^{sg} \right)^{-2} \frac{\dab}{8} 
       \displaystyle \exp\left(\frac{\dab^2}{8} \right) \left[ K_1\left(\frac{\dab^2}{8} \right) - K_0\left(\frac{\dab^2}{8} \right) \right]
\end{equation}
which is the SG force kernel of a thin Gaussian-stratified disc.

\section{Formulation as a convolution product}\label{app: convolution product}

The formulation of the SG forces as a convolution product relies on rewriting the Bessel Kernel, and ultimately the normalized distance $d^2_{ab}$, in terms of shifted arguments.
Starting with the expression of $d^2_{ab}$:
\begin{equation}
\begin{array}{lll}
d^2_{ab} & = & \displaystyle 2 \frac{r^2+r'^2-2r r' \cos{(\varphi-\varphi')}}{H^2_a(r', \varphi') + H^2_b(r, \varphi)} 
\end{array}
\end{equation}
We can rewrite it as:
\begin{equation}
\begin{array}{lll}
d^2_{ab} & = & \displaystyle 2 \frac{r/r'+r'/r-2 \cos{(\varphi-\varphi')}}{H^2_a(r', \varphi')/(r r') + H^2_b(r, \varphi) / (r r') }
\end{array}
\end{equation}
By substituting $r=e^u$ and $r'=e^{u'}$ in the numerator, we obtain the correct form for the convolution. 
However, the denominator requires additional constraints: $H^2_a(r', \varphi')/(r r')$ and $H^2_b(r, \varphi) / (r r')$ must be solely functions of $r'/r$, which occurs only if the scale heights are linear with the polar radius.
Under this constraint, the normalized distance is expressed as:
\begin{equation}
\begin{array}{lll}
d^2_{ab} & = & \displaystyle 2 \frac{e^{u-u'} + e^{-(u-u')} - 2 \cos{(\varphi-\varphi')} }{h^2_a e^{u'-u} + h^2_b e^{u-u'}} \\ [12pt]
         & = & \displaystyle 4 \frac{\cosh{(u-u')} - \cos{(\varphi-\varphi')} }{h^2_a e^{u'-u} + h^2_b e^{u-u'}}
\end{array}
\end{equation}
where $h_a=H_a(r')/r'=const.$ and $h_b=H_b(r)/r=const$.

The radial and azimuthal components of the Bessel kernel, expressed in a form suitable for convolution on a logarithmic radial grid, are given by:
\begin{equation}
\left\{
\begin{array}{lll}
\text{K}_r(X, \varphi)  & = & \displaystyle \frac{L_{ab}(\dab)}{\pi \dab^2} \left[ \frac{e^{- X}}{h^2_{ab}} \right]^{3/2} \left[ e^{X} - \cos{(\varphi)} \right]   \\ [12pt]
\text{K}_\varphi(X, \varphi)  & = & \displaystyle \frac{L_{ab}(\dab)}{\pi \dab^2} \left[ \frac{e^{- X}}{h^2_{ab}} \right]^{3/2} \sin{(\varphi)}
\end{array}
\right.
\end{equation}
where $X=u-u'$.
The square of the geometric mean, the square of the normalized distance and the auxiliary function $L_{ab}$ are given by:
\begin{equation}
\begin{array}{lll}
h^2_{ab}     & = & \Hab/(rr') =  \displaystyle \frac{h^2_a e^{-X} + h^2_b e^{X}}{2} \\ [12pt]
d^2_{ab} & = & \displaystyle 2 \frac{\cosh{X} - \cos{\varphi} }{h^2_{ab}} \\ [12pt]
L_{ab}(\dab) & = &  \displaystyle \sqrt{\pi} \frac{\dab^2}{8}  \exp\left(\frac{\dab^2}{8} \right) \left[ K_1\left(\frac{\dab^2}{8} \right) - K_0\left(\frac{\dab^2}{8} \right) \right] 

\end{array}
\end{equation}
Note that the expression for $L_{ab}$ differs slightly from the standard form of the Bessel kernel presented in Eq. \ref{Eq: SG force kernel exact}.

\section{Limitations of the full spectral method for non-constant disc aspect ratios}\label{app: limitations of the full spectral method}

\begin{figure}[h]
\centering
\includegraphics[width=\hsize]{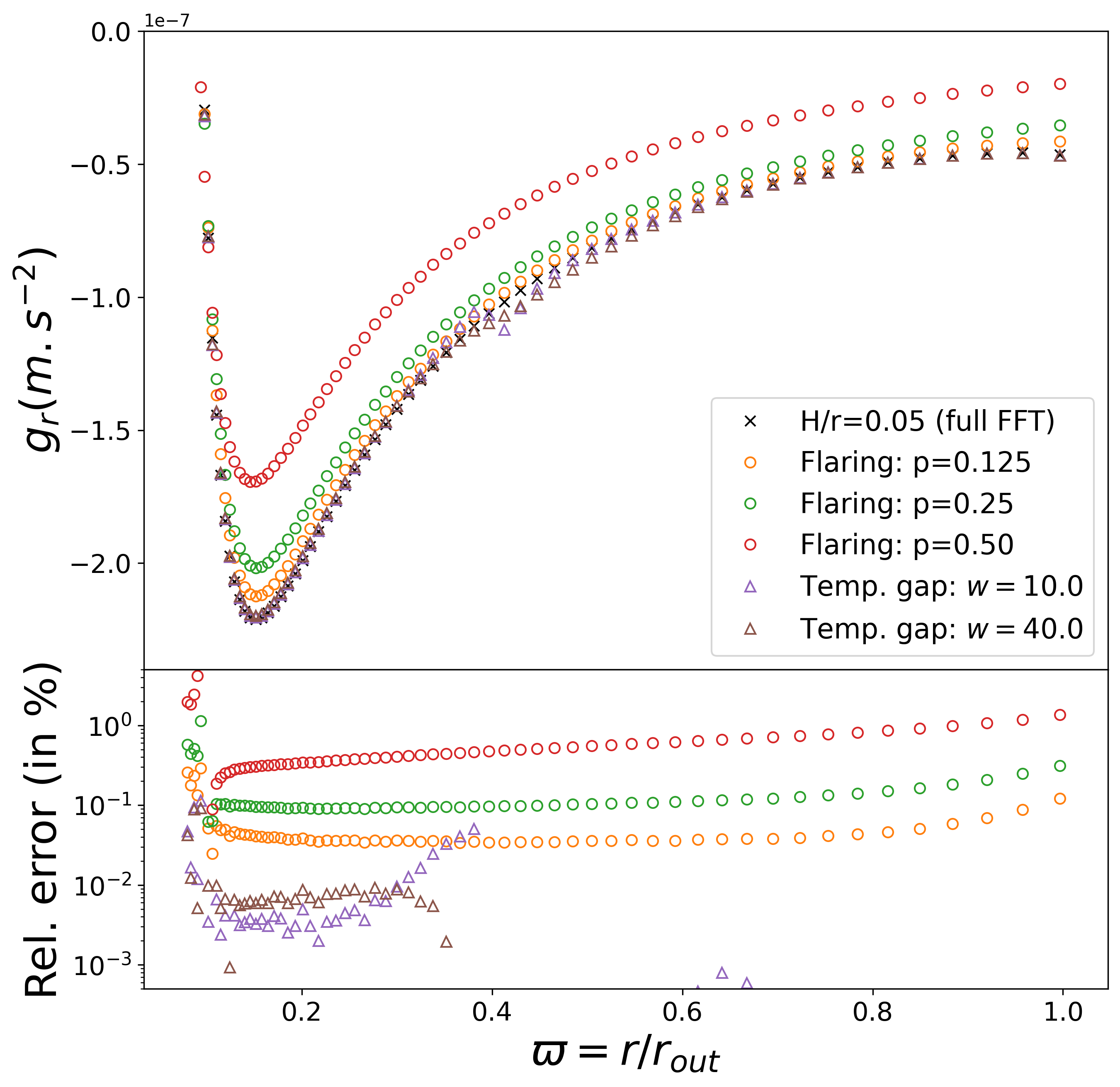}

\caption{Radial SG force of the power law disc for different disc flarings and when a temperature gap is introduced.
} 
\label{fig: compare FFT and semi-FFT}
\end{figure}

The high performance of the full spectral method, makes necessary to employ it despite not fully satisfying its application hypothesis, namely the constant $h=H/r$ condition. 
To quantify the error arising from deviating from this condition, we compared a constant flaring disc setup to reference setups featuring radial power law flaring and local temperature gaps.
For the reference setups, we used the semi-spectral method implemented in \NIRVANA, as detailed in Sect. \ref{sec: efficient numerical method}.
Our numerical setup is:
\begin{equation}
\left\{
\begin{array}{lll}
\Sigma_0(r)        &=& 20000 \, (r/1\,\mbox{AU})^{-1.5} \, \mbox{g.cm}^{-2}  \\
r_0                &=& 100 \, \text{AU} \\
(r_{in}, r_{out})  &=& (20, 250) \, \mbox{AU} \\
(N_r, N_\varphi)    &=& (1200, 2800) \\
\end{array}
\right.
\end{equation}
To investigate the impact of flaring, we choose  
\begin{equation}
H/r  = 0.05 \, (r/1\,\mbox{AU})^{p}
\end{equation}
with $p \in [0.125, 0.25, 0.50]$.
For the second setup, testing the impact of a temperature gap, we used 
\begin{equation}
H/r = 0.05 \, \sqrt{1-0.95 \, \exp\left(-\left(\frac{r-100 \mbox{AU}}{w}\right)^2 \right) }
\end{equation}
with $w \in [10, 40] \, \mbox{AU}$.

Our results, presented in Figure \ref{fig: compare FFT and semi-FFT}, demonstrate that as the $h=const.$ condition deviates, the error increases across the entire disc.
The error ranged from approximately a few percent for $p=0.125$, to a few tens of percent for $p=[0.25, 0.5]$.
In contrast, the impact of the gap was localized and minimal, with the worst-case error being of the order of 10\%. 

We conclude that using a full spectral method remains relatively accurate for disc flarings not exceeding 0.125, and that local temperature variations do not significantly impact the error.
Therefore, for production runs of GI, the highest concern should not be transient structures like spirals or clumps but rather the global disc flaring. 
If the full FFT method is to be used on setups where the disc flaring is not zero, it may be wiser to simply choose a higher disc aspect ratio $h$ when used in the numerical method for computing SG.

\section{Indirect term}\label{app: indirect term}

The indirect term associated to the pull exerted by fluid `a' on the star is:
\begin{equation}\label{Eq: indirect term proof}
\begin{array}{ll}
\Phi_{a,\text{ind}} & = -\vec{a}_{\odot} \cdot \vr  \\
                  & = \displaystyle \iint\limits_{disc} \int\limits_{z=-\infty}^{\infty}  
                      \frac{\Sigma_a(r') e^{-\frac{1}{2} \left(z'/H_a^{sg} \right)^2}}{\sqrt{2 \pi} H_a^{sg}(r')} 
                      \frac{\left( r' \Vec{e}_{r'} + z' \vec{e}_{z'} \right)}{\left( r'^2 + z'^2 \right)^{3/2}}  \cdot \vr \, d^2 \vr' \\
                  & = \displaystyle \iint\limits_{disc} 
                       \frac{\Sigma_a(r')}{\sqrt{2 \pi}} r r' \cos{\left(\varphi-\varphi' \right)} (H_a^{sg})^{-3} 
                       F_1\left( d_a^2 \right) \, d^2 \vr'
\end{array}
\end{equation}
where $d_a=||\vr'||/H_a^{sg}$ and $F_1$ is the function defined by Eq.~\ref{Eq: function Fn}.
Using the results of appendix~\ref{app:useful function}, we get:
\begin{equation}
\Phi_{a, \text{ind}} (\vr) = \sqrt{2} r \iint\limits_{disc} \Sigma_a(\vr')  \cos{\left(\varphi-\varphi'\right)} \text{K}_a(d_a) d^2 \vr' 
\end{equation}
where $\displaystyle \text{K}_a(d_a)=\frac{1}{\sqrt{\pi}} (H_a^{sg})^{-2} \frac{d_a}{4} \exp\left(\frac{d_a^2}{4} \right) \left[ K_1\left(\frac{d_a^2}{4} \right) - K_0\left(\frac{d_a^2}{4} \right) \right]$.
We emphasise that thus far, we have not used the vertical integral component of our reasoning. Indeed, it is worth noting that the vertical integration employed in Eq.~\ref{Eq: indirect term proof} is inherently encompassed within the definition of the 3D indirect term potential.
Thus, to align this indirect term with our approach, we must calculate the following quantity:
\begin{equation}
\begin{array}{ll}
- \int\limits_{z=-\infty}^{\infty} \rho_a(\vr,z) \Vec{\nabla}_{3D} \Phi_{a, \text{ind}} \, dz 
                & = - \left[ \int\limits_{z=-\infty}^{\infty} \rho_a(\vr,z) \, dz \right] \Vec{\nabla}_{2D} \Phi_{a, \text{ind}} \\ [8pt]
                & = - \Sigma_a(\vr) \Vec{\nabla}_{2D} \Phi_{a, \text{ind}} \\
\end{array}
\end{equation}
We finally obtain the gradient of the indirect potential:
\begin{equation}
\begin{array}{ll}
\Vec{\nabla}_{2D} \Phi_{a, \text{ind}} & = \sqrt{2} \iint\limits_{disc} \Sigma_a(\vr') \text{K}_a(\vr')  \\ 
& \left[ \cos{\left(\varphi-\varphi'\right)} \, \vec{e}_r -  \sin{\left(\varphi-\varphi'\right)} \, \vec{e}_\varphi \right]  d^2 \vr' \\ [10pt]
                                       & = \left(
\begin{array}{rcc}
   A_{disc} \cos{\left(\varphi\right)} & + & B_{disc} \sin{\left(\varphi\right)} \\
  -A_{disc} \sin{\left(\varphi\right)} & + & B_{disc} \cos{\left(\varphi\right)}
\end{array}
\right)
\end{array}
\end{equation}
with $A_{disc}=\sqrt{2} \iint\limits_{disc} \Sigma_a(\vr') \text{K}_a(\vr') \cos{\left( \varphi'\right)} d^2 \vr' $ and \\ $B_{disc}=\sqrt{2} \iint\limits_{disc} \Sigma_a(\vr') \text{K}_a(\vr') \sin{\left( \varphi'\right)} d^2 \vr'$.

\section{Conservative force}\label{app: conservative force}

When the scale heights $H_a^{sg}$ and $H_b^{sg}$ are space constants, we can derive the expression for the 2D gravitational potential, as we will demonstrate next.
With the help of Eq.~\ref{Eq: grav. field 2D general}, it is possible to reformulate the expression for the 2D gravitational field as the gradient of a potential: 
\begin{equation}
\begin{array}{lll}
\Vec{\mathcal{G}}_{2D}^{a\rightarrow b} &=& \displaystyle - \Vec{\Nabla}_{2D} \left[ \, \,  \int\limits_{z=-\infty}^{\infty} \frac{e^{-\frac{1}{2}(z/H_b^{sg})^2}}{\sqrt{2 \pi} H_{b}^{sg}}  \Phi_a(\vr,z) \, dz \right] \\ [10pt]
                                        &=& - \Vec{\Nabla}_{2D} \Phi_{ab} (\vr)                                        
\end{array}
\end{equation}
The combined potential, $\Phi_{ab}$, can be rearranged by employing the integral formulation for $\Phi_a$:
\begin{equation}
\begin{array}{lll}
\Phi_{ab}(\vr) & = & - \displaystyle \frac{1}{2\pi} \frac{1}{H_a^{sg} H_b^{sg}} \iint\limits_{disc} \Sigma_a(\vr') \\
          &   & \left( \, \, \iint\limits_{z, z'=-\infty}^{+\infty} 
                   \displaystyle \frac{e^{-\frac{1}{2} \left(z/H_{b}^{sg}\right)^2} e^{-\frac{1}{2} \left(z'/H_{a}^{sg}\right)^2}}{\left(s^2+(z-z')^2\right)^{{1}/{2}}} \, dz \, dz' \right)  d^2\vr'
\end{array}
\end{equation}
The integral expression above can be expressed in a closed form by following, and adapting, the steps outlined in appendices~\ref{app:useful function}-\ref{app: integration}.
Ultimately, we obtain:
\begin{equation}
\begin{array}{lll}
\Phi_{ab}(\vr) & = &  - \displaystyle \iint\limits_{disc}  \frac{\Sigma_a(\vr')}{2 \sqrt{\pi} \Hab} \exp\left( \frac{\dab^2}{8} \right) K_0 \left(\frac{\dab^2}{8} \right) d^2\vr' 
\end{array}
\end{equation}
Above quantity represents the 2D gravitational potential of a bi-fluid system, where both fluids exhibit a Gaussian vertical stratification.

\end{appendix}

\label{LastPage}

\end{document}